\documentclass[12pt,letterpaper]{article}
\usepackage[usenames, dvipsnames]{xcolor}
\usepackage{jcapmod}

\usepackage{tocloft}
\usepackage[]{todonotes}
\usepackage{verbatim}
\usepackage[utf8]{inputenc}
\usepackage{mathrsfs}
\usepackage{cleveref}
\usepackage[shortlabels]{enumitem}
\usepackage{setspace,caption}
\usetikzlibrary{external}
\makeatletter
\newsavebox\myboxA
\newsavebox\myboxB
\newlength\mylenA

\definecolor{cornellRed}{HTML}{B31B1B}
\definecolor{cornellBlue}{HTML}{0068AC}
\definecolor{cornellGreen}{HTML}{6EB43F}

\newtheorem{theorem}{Theorem}

\newtheorem{definition}{Definition}

\usepackage{bbm} 					
\usepackage{slashed} 				
\usepackage{graphicx}				
\usepackage{subcaption}			
\usepackage{psfrag}				
\usepackage{tensor}				
\usepackage{fouridx}				
\usepackage{bm}					
\usepackage{mdframed}				
\usepackage{multirow}				
\usepackage{soul}					
\usepackage{bbold}				
\usepackage{multicol}				

\usepackage{amsmath}
\usepackage{amssymb}
\usepackage{amsfonts}
\usepackage{mathtools}

\usepackage{feynmf}
\usepackage{marvosym}

\usepackage{import}
\usepackage[ruled,vlined]{algorithm2e}

\usepackage{multirow}
\usepackage{tabularx}
\usepackage{ragged2e}
\usepackage{booktabs}
\newcolumntype{Y}{>{\centering\arraybackslash}X}

\AtBeginDocument{
	\heavyrulewidth=.08em
	\lightrulewidth=.05em
	\cmidrulewidth=.03em
	\belowrulesep=.65ex
	\belowbottomsep=0pt
	\aboverulesep=.4ex
	\abovetopsep=0pt
	\cmidrulesep=\doublerulesep
	\cmidrulekern=.5em
	\defaultaddspace=.5em
}

\newcommand*\xoverline[2][0.75]{%
    \sbox{\myboxA}{$\m@th#2$}%
    \setbox\myboxB\null
    \ht\myboxB=\ht\myboxA%
    \dp\myboxB=\dp\myboxA%
    \wd\myboxB=#1\wd\myboxA
    \sbox\myboxB{$\m@th\overline{\copy\myboxB}$}
    \setlength\mylenA{\the\wd\myboxA}
    \addtolength\mylenA{-\the\wd\myboxB}%
    \ifdim\wd\myboxB<\wd\myboxA%
       \rlap{\hskip 0.5\mylenA\usebox\myboxB}{\usebox\myboxA}%
    \else
        \hskip -0.5\mylenA\rlap{\usebox\myboxA}{\hskip 0.5\mylenA\usebox\myboxB}%
    \fi}
\makeatother

\newcommand{\cA}{\mathcal{A}}
\newcommand{\cC}{\mathcal{C}}

\newcommand{\cK}{\mathcal{K}}

\newcommand{\cM}{\mathcal{M}}

\newcommand{\cO}{\mathcal{O}}

\newcommand{\cT}{\mathcal{T}}

\newcommand{\cV}{\mathcal{V}}

\newcommand{\ZZ}{\mathbb{Z}}

\definecolor{cobalt}{RGB}{44, 98, 120}
\definecolor{celadon}{rgb}{0.67, 0.88, 0.69}
\definecolor{dm}{cmyk}{.20, 0, .30, 0}
\definecolor{burgundy}{rgb}{0.5, 0.0, 0.13}
\definecolor{plotBlue}{RGB}{94, 130, 181}

\newcommand{\R}{\mathbb{R}}

\hypersetup{
  colorlinks,
  citecolor=Violet,
  linkcolor=cobalt,
  urlcolor=Blue}

\DeclareSymbolFontAlphabet{\mathbb}{AMSb}

\newif\iffastcompile

\fastcompilefalse

\iffastcompile
\newcommand{\cl}[1]{}
\newcommand{\lm}[1]{}
\newcommand{\md}[1]{}
\newcommand{\ab}[1]{}
\newcommand{\art}[1]{}
\else
\newcommand{\cl}[1]{\todo[color=burgundy!30, size=\scriptsize, bordercolor=burgundy!30]{CL: #1}}
\newcommand{\lm}[1]{\textcolor{red}{[#1]}}
\newcommand{\md}[1]{\textcolor{blue}{[#1]}}
\newcommand{\art}[1]{\textcolor{green}{[#1]}}
\fi

\newcommand{\email}[1]{\href{mailto:#1}{#1}}

\ProvideTextCommandDefault{\Dbar}{%
\leavevmode\lower.5ex\rlap{\hskip-.07em\accent"16}D%
}

\begin{document}
	\newcommand{\main}{.}
\begin{titlepage}

\setcounter{page}{1} \baselineskip=15.5pt \thispagestyle{empty}
\setcounter{tocdepth}{2}

\bigskip\

\vspace{1cm}
\begin{center}
{\fontsize{19}{26} \bfseries Bounding the Kreuzer-Skarke Landscape}

\end{center}

\vspace{0.45cm}

\begin{center}
\scalebox{0.95}[0.95]{{\fontsize{14}{30}\selectfont  Mehmet Demirtas, Liam McAllister, and Andres Rios-Tascon}}

\end{center}

\begin{center}

\textsl{Department of Physics, Cornell University, Ithaca, NY 14853, USA}\\

\vspace{0.25cm}

\email{ md775@cornell.edu, mcallister@cornell.edu, ar2285@cornell.edu }
\end{center}

\vspace{0.6cm}
\noindent

We study Calabi-Yau threefolds with large Hodge numbers by constructing and counting triangulations of reflexive polytopes. By counting points in the associated secondary polytopes, we show that the number of fine, regular, star triangulations of polytopes in the Kreuzer-Skarke list is bounded above by $\binom{14{,}111}{494} \approx 10^{928}$.  Adapting a result of Anclin on triangulations of lattice polygons, we obtain a bound on the number of triangulations of each 2-face of each polytope in the list.  In this way we prove that the number of topologically inequivalent Calabi-Yau hypersurfaces arising from the Kreuzer-Skarke list is bounded above by $10^{428}$.  We introduce efficient algorithms for constructing representative ensembles of Calabi-Yau hypersurfaces, including the extremal case $h^{1,1}=491$, and we study the distributions of topological and physical data therein.  Finally, we demonstrate that neural networks can accurately predict these data once the triangulation is encoded in terms of the secondary polytope.

\noindent
\vspace{2.1cm}

\noindent\today

\end{titlepage}
\tableofcontents\newpage

\section{Introduction}

String theory gives rise to an immense wilderness of four-dimensional effective theories.  A first step to understanding these theories is to chart out the regions in theory space that are populated by specific classes of solutions.  Compactifications of critical string theories on Calabi-Yau threefolds, or on orientifolds thereof, are perhaps the best-understood class for this purpose, because of the perspective afforded by algebraic geometry.

Although Calabi-Yau threefolds are not fully classified, or even known to be finite in number, one can learn much by studying subclasses that enjoy additional structure.  Calabi-Yau threefolds that are hypersurfaces in toric varieties can be treated combinatorially because the ambient toric varieties correspond to certain triangulations of reflexive polytopes in $\mathbb{Z}^4$, as we will recall below.
Kreuzer and Skarke famously enumerated the 473{,}800{,}776 distinct four-dimensional reflexive polytopes \cite{Kreuzer:2000xy}, and their list has served as a wellspring of topological data of Calabi-Yau threefolds (see e.g.~\cite{Altman:2014bfa,Huang:2018gpl}).  However, the actual number of inequivalent Calabi-Yau threefolds arising as hypersurfaces in toric varieties has not been known, even to the nearest thousand orders of magnitude.

In this work we prove an upper bound, $N_{\text{CY}} < 1.65 \times 10^{428}$, on the number of inequivalent Calabi-Yau threefold hypersurfaces in toric varieties.  We also show that the number of fine, regular, star triangulations (FRSTs) of  four-dimensional reflexive polytopes is bounded above by $N_{\text{FRST}} < 1.53 \times 10^{928}$.  Both upper bounds are dominated by the single polytope with the largest number of points: this is a four-simplex containing 496 lattice points, which engenders Calabi-Yau threefolds with $h^{1,1}=491$ and $h^{2,1}=11$.

To grapple with these astronomically large discrete sets, we then develop novel algorithms for fair sampling of triangulations and of Calabi-Yau threefolds.  Enumerating FRSTs of polytopes with hundreds of points and then operating on the triangulation data is computationally demanding, and to reach acceptable speeds we were obliged to adapt and combine a considerable number of software packages and libraries.
Our methods allow rapid enumeration of triangulations of even the largest polytope in the Kreuzer-Skarke list, and give samples that are representative in the range of $h^{1,1}$ where we are able to test them.

Finally, the immense size of discrete data sets in string theory, such as those encountered here, suggests the use of data science techniques --- see \cite{2004JHEP...07..069A,2014JHEP...08..010A,2016JHEP...03..045C,2017PhLB..774..564H,Krefl:2017yox,2017JHEP...08..038R,2017JHEP...09..157C,2018arXiv180503615A,2018PhLB..785...65B,2019ForPh..67l0084C,2018arXiv180902547K,2018arXiv180902612E,2019NuPhB.940..113M,2018arXiv181206960C,2019PhLB..795..700B,2019JHEP...06..003H,2019PhLB..79834889H,2019arXiv191008605A,2020ForPh..68e0005H,2020arXiv200313339D,2020arXiv200616619H,2020arXiv200700009B} as well as the reviews \cite{RUEHLE20201} and \cite{2018arXiv181202893H}.  In the final part of this work, we develop machine learning algorithms to predict the topological data of Calabi-Yau threefolds.  We demonstrate that these are orders of magnitude faster than our enumerative algorithms, have outstanding accuracy at modest Hodge numbers, and remain accurate for some quantities even at $h^{1,1}=491$.  The key idea is to represent the data of a regular triangulation in terms of the \emph{GKZ vector} --- introduced in Definition \ref{def:gkz_vector} below --- that arises in the secondary polytope construction of Gelfand, Kapranov, and Zelevinsky (GKZ) \cite{Gelfand1994}.

The organization of this paper is as follows.  In \S\ref{sec:reg_traings_and_sec_poly} we review key elements of Calabi-Yau threefold hypersurfaces in toric varieties, and of the secondary polytope of a point configuration.
In \S\ref{fullsec:bound_lat_poly} we derive upper bounds on the number of FRSTs and the number of distinct Calabi-Yau threefolds that result from polytopes in the Kreuzer-Skarke list.
In \S\ref{fullsec:sampling} we present algorithms for fair sampling of FRSTs and Calabi-Yau threefolds, as well as our software implementation (in \S\ref{sec:software}).
In \S\ref{sec:ml} we demonstrate that deep neural networks can very accurately predict topological data of Calabi-Yau threefold hypersurfaces.
We conclude in \S\ref{conclusions}.
Appendix \S\ref{sec:app_randomsample} contains the results of testing our sampling algorithms.

\section{Regular Triangulations and the Secondary Polytope}\label{sec:reg_traings_and_sec_poly}
We start by briefly reviewing Batyrev's construction \cite{Batyrev:1994hm} of Calabi-Yau threefolds as hypersurfaces in toric varieties. Let $\Delta$ and $\Delta^\circ$ be a dual pair of four-dimensional reflexive polytopes, and let $\mathcal{T}$ be a triangulation of $\Delta^{\circ}$. When $\mathcal{T}$ is a \textit{fine, regular, star} triangulation (FRST), conditions that we will presently discuss, it defines a subdivision of the normal fan of $\Delta$, and so defines a four-dimensional toric variety whose generic anticanonical hypersurface is a smooth Calabi-Yau threefold.

Each condition on the triangulation $\mathcal{T}$ plays an important role in the construction. $\mathcal{T}$ is called \textit{star} if all of its full-dimensional simplices have the origin as a vertex. This allows one to define cones forming a complete fan, and hence construct a compact toric variety. $\mathcal{T}$ is called \textit{fine} if every point in the point configuration, which in our case consists of all lattice points not interior to facets,\footnote{Points strictly interior to facets correspond to divisors of the ambient variety that do not intersect the Calabi Yau hypersurface transversely. See e.g.~Appendix B of \cite{Braun:2017nhi} for a detailed discussion.} is a vertex of a simplex in $\mathcal{T}$. This ensures that the singularities in the toric variety are sufficiently mild so that the Calabi-Yau is smooth. Finally, $\mathcal{T}$ is called \textit{regular} if it can be obtained as the projection of the lower faces of the convex hull of a point set in one higher dimension. This is required so that the toric variety is projective, and hence K\"ahler --- the hypersurface then inherits this K\"ahler structure.

\begin{figure}[!ht]
    \centering
    \includegraphics[width=\textwidth]{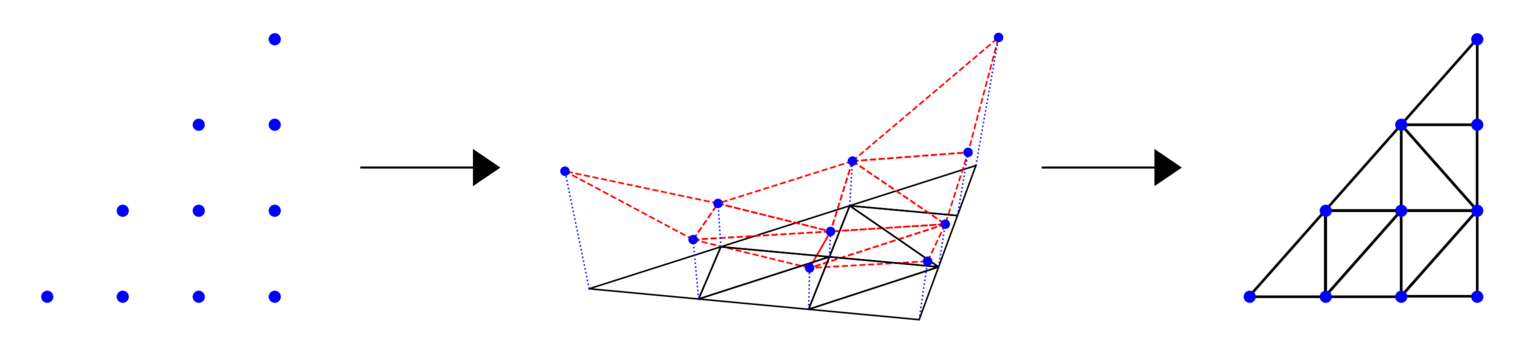}
	\caption{Regular triangulations descend from projections of the lower faces of a convex hull in one higher dimension. In the figure, a two-dimensional point set is lifted by a set of heights into a three-dimensional convex hull. The lower faces (shown in red) induce a triangulation on the original point set via their projection.}
	\label{fig:reg_triang}
\end{figure}

Let us now dissect the regularity condition, which will be crucial in our analysis. A triangulation $\mathcal{T}$ of a configuration $\mathcal{A}$ of $n$ points $p_i \in \mathbb{R}^d$, $i=1,\ldots,n$, is regular if it can be constructed as follows. One starts with the points of $\mathcal{A}$ and lifts them into one higher dimension, i.e.~into $\mathbb{R}^{d+1}$, by adding an extra coordinate and defining heights $h_i \in \mathbb{R}$, $i=1,\ldots,n$. One can then construct the convex hull. For generic heights, the lower faces of this convex hull will be simplices that can be projected down to produce a triangulation of $\mathcal{A}$. A schematic of this process is shown in Figure~\ref{fig:reg_triang}. If one of the lifted points, $p$, lies in the interior of the convex hull, then the resulting triangulation does not include $p$. Note that within the space of heights there are lower-dimensional subspaces that do not result in triangulations of $\mathcal{A}$, but rather in regular subdivisions that do not fully divide $\mathcal{A}$ into simplices.

The regular triangulations of a point configuration have a very rich structure studied by Gelfand, Kapranov, and Zelevinsky (GKZ) \cite{Gelfand1989,Gelfand1991}. An extensive discussion of their results can be found in their book \cite{Gelfand1994} and in the more recent textbook \cite{DeLoera2010}, which also provides more general information on triangulations.  Here we recall the constructions and theorems that will be necessary for our purposes. The main concepts we will need are those of the secondary polytope and secondary fan of a $d$-dimensional point configuration $\mathcal{A}$ with $n$ points. To this end, we first define the GKZ vector of a triangulation.

\begin{definition}\label{def:gkz_vector}
	\emph{(GKZ Vector)} Let $\mathcal{A}$ be a point configuration with points $p_i\in\R^d$, $i=1,\ldots,n$, and let $\mathcal{T}$ be a triangulation of $\mathcal{A}$. We define a map $\varphi_\mathcal{A}:\{\mathrm{triangulations\ of\ }\mathcal{A}\}\rightarrow \R^n$ such that $\varphi_{\mathcal{A}}(\mathcal{T})$ is a vector whose $i$th component is given by
	\begin{equation}
		\varphi^i_\mathcal{A}(\mathcal{T}):=\sum_{\{\sigma\in\mathcal{T}|p_i\in\mathrm{vert}(\sigma)\}}\mathrm{vol}(\sigma)\,,
	\end{equation}
	where the sum is over all maximal simplices $\sigma$ that have $p_i$ as a vertex. This is called the \emph{GKZ vector} (or \emph{volume vector}) of $\mathcal{T}$.
\end{definition}

The secondary polytope can now be constructed from the GKZ vectors as follows.

\begin{definition}
	\emph{(Secondary Polytope)} Let $\mathcal{A}$ be a point configuration. The \emph{secondary polytope} $\Sigma(\mathcal{A})$ is the convex hull of the GKZ vectors associated to all the triangulations of $\mathcal{A}$,
	\begin{equation}
		\Sigma(\mathcal{A}):=\mathrm{conv}\Bigl\{\varphi_\mathcal{A}(\mathcal{T})|\mathcal{T}\mathrm{\ is\ a\ triangulation\ of\ }\mathcal{A}\Bigr\}\,.
	\end{equation}
\end{definition}
We adopt the convention that the standard simplex has unit volume, which implies that if the points in $\mathcal{A}$ are lattice points, as in our situation, then the secondary polytope is a lattice polytope in $\ZZ^n$. The key result due to GKZ is the following theorem.
	
\begin{theorem}
\emph{(Gelfand, Kapranov, Zelevinsky 1989)} For a point configuration $\mathcal{A}$, the vertices of the secondary polytope $\Sigma(\mathcal{A})$ are in one-to-one correspondence with the regular triangulations of $\mathcal{A}$.\footnote{In fact, the theorem is much stronger: it asserts that the face lattice of $\Sigma(\mathcal{A})$ is isomorphic to the refinement poset of regular polyhedral subdivisions of $\mathcal{A}$. For more details see \cite{Gelfand1994,DeLoera2010}.}
\end{theorem}

Let us now relate the secondary polytope $\Sigma(\cA)$ to the definition of regularity described above. Given a point configuration $\cA$ with $n$ points, the set of height vectors that result in a given triangulation forms (the interior of) a full-dimensional cone in $\R^n$. The set of such cones (including their boundaries and intersections) from all of the regular triangulations forms a complete fan, called the \emph{secondary fan}, which can be equivalently defined as follows \cite{Gelfand1989,Gelfand1991}:

\begin{definition}
	\emph{(Secondary Fan)} The \emph{secondary fan} of a point configuration $\mathcal{A}$ is the normal fan of the secondary polytope $\Sigma(\mathcal{A})$.
\end{definition}

Since the construction of smooth Calabi-Yau hypersurfaces requires triangulations to be fine and star, we need to determine the regions in the secondary fan that correspond to FRSTs. To this end, we make use of the following theorem \cite{DeLoera2010}:

\begin{theorem}\label{theorem:height_paths}
    Let $\cA$ be a point configuration. Let $\mathbf{h}$ and $\mathbf{h}'$ be the generic height vectors defining two regular triangulations $\cT$ and $\cT'$. Finally, let $\cT_t$ be the triangulation (or subdivision) obtained from the height vector $\mathbf{h}_t:=(1-t)\mathbf{h}+t\mathbf{h}'$ for $t\in[0,1]$. Then:
    \begin{enumerate}[a)]
        \item All subdivisions $\cT_t$ are either triangulations or bistellar flips between two triangulations.
        \item All subdivisions $\cT_t$ contain all the simplices that are common to both $\cT$ and $\cT'$.\label{theorem:height_paths_connectedness}
    \end{enumerate}
\end{theorem}

Theorem~\ref{theorem:height_paths} implies that the cones from FRSTs form a \textit{subfan with convex support} of the secondary fan.\footnote{We remark that the cones in the secondary fan are closely related to the K\"ahler cones of the ambient varieties: see \cite{oda1991}.} To see this we use the simple fact that a triangulation is fine and star if and only if for each point other than the origin there is a one-simplex having this point and the origin as its vertices.

\begin{figure}[!ht]
    \centering
	\includegraphics[width=\textwidth]{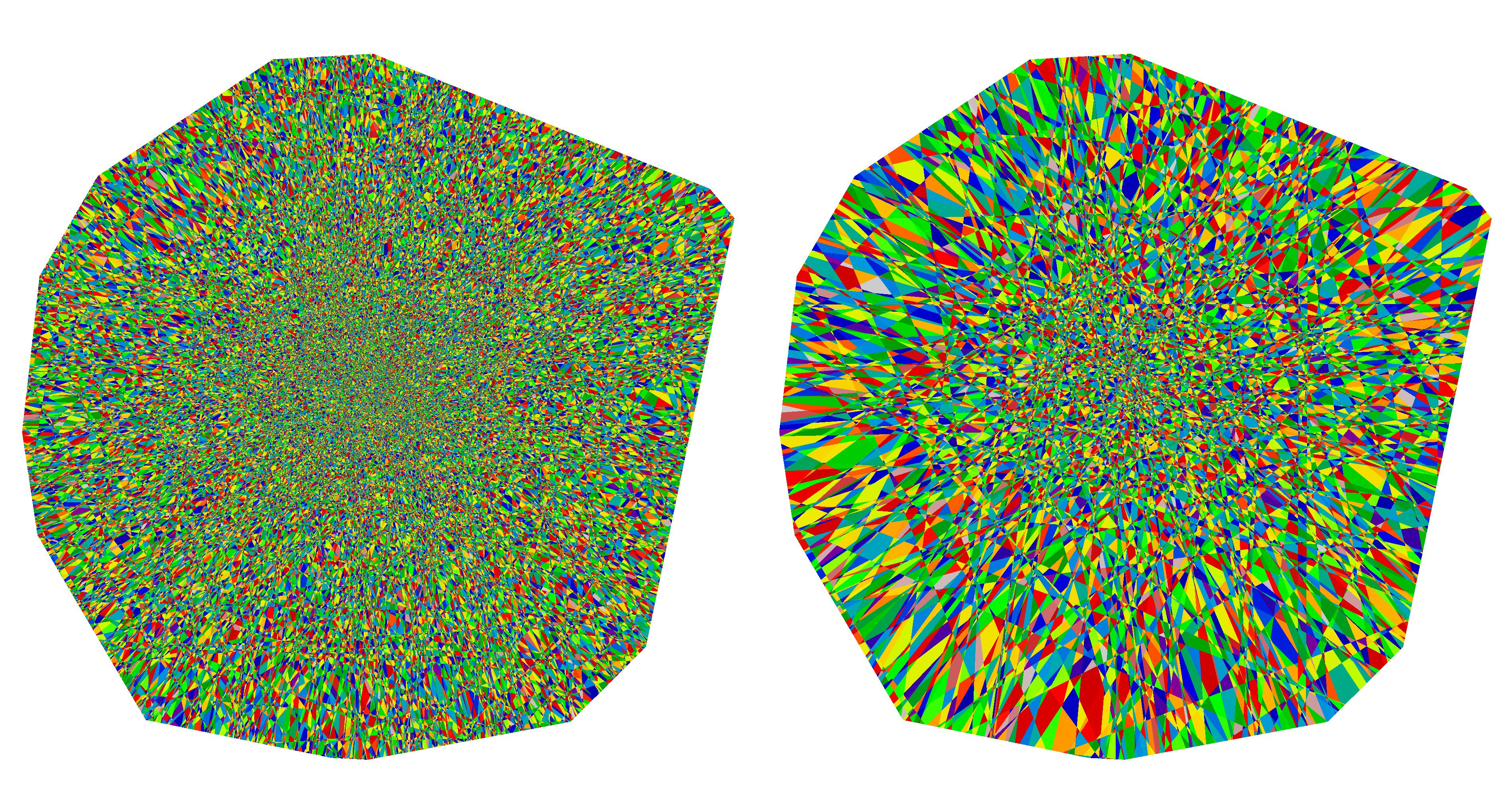}
	\caption{A two-dimensional cross section of the secondary subfan of FRSTs of the largest four-dimensional reflexive polytope (left), and the result of combining regions therein that give rise to the same Calabi-Yau (right). Each colored region on the left (resp. right) corresponds to a different FRST (resp. Calabi-Yau), and the outermost boundaries are defined by heights beyond which the triangulation becomes non-fine or non-star.}
	\label{fig:491_xsec}
\end{figure}

A two-dimensional cross section of the subfan of FRSTs of the largest reflexive polytope in the Kreuzer-Skarke database is shown in Figure~\ref{fig:491_xsec}.

\section{Upper Bounds on the Number of Triangulations}\label{fullsec:bound_lat_poly}

As a step toward understanding the statistics of string vacua, it would be valuable to compute the number $N_{\mathrm{CY}}$ of distinct Calabi-Yau hypersurfaces arising from the Kreuzer-Skarke list.  In this work we will derive a rigorous \textit{upper bound} on $N_{\mathrm{CY}}$, leaving a direct computation of $N_{\mathrm{CY}}$ as a challenge for future work.

In this section, we describe a general procedure for deriving upper bounds on the number of regular triangulations of lattice polytopes. To our knowledge, our result in equation \eqref{eqn:bound_reg_triang} is the best upper bound available in the literature for the number of regular triangulations of lattice polytopes\footnote{Lattice polygons have been studied more extensively, see e.g.~\cite{Anclin2003,Kaibel03,DeLoera2010}.} in $d\geq 3$. We then specialize to the case of fine, regular, star triangulations of reflexive polytopes, and derive an upper bound for the number of topologically distinct Calabi-Yau threefolds that can arise.

\subsection{Bounding the number of regular triangulations of lattice polytopes}\label{sec:bound_lat_poly}

Our derivation of an upper bound relies on the construction of the secondary polytope described in \S\ref{sec:reg_traings_and_sec_poly}. In particular, we will make crucial use of the fact that the vertices of the secondary polytope are in bijective correspondence with regular triangulations of the original point configuration.

Consider a set $\mathcal{A}$ of $n$ points in $\ZZ^d$ whose convex hull is a $d$-dimensional polytope.\footnote{The set of $n$ points need not include all the lattice points contained in the convex hull.}  Let $V$ be the volume of the polytope, in units where the standard simplex has unit volume. Definition~\ref{def:gkz_vector} then implies that the components $\varphi_{\mathcal{A}}^{i}$ of the GKZ vectors take integer values ranging from $0$ to $V$.
By counting the lattice points in the box defined by $0 \le \varphi_{\mathcal{A}}^{i} \le V$, $i=1,\ldots,n$, we obtain an upper bound on the number of regular triangulations of $\mathcal{A}$.

We can obtain a tighter bound by using the fact that the sum of all components of each GKZ vector must be $(d+1)V$, which follows from the fact that each simplex has $d+1$ vertices and hence its volume is counted $d+1$ times.  The counting problem then reduces to finding the number of solutions to the equation
\begin{equation}
	\sum_{i=1}^{n}x_i=(d+1)V\,,\ \mathrm{with\ }x_i\in\mathbb{Z}_{\geq 0}\,.
\end{equation}
This is a straightforward combinatorial problem that yields an upper bound of
\begin{equation}
    N_\mathrm{reg. triang.}\leq\binom{(d+1)V+n-1}{n-1}\,.\label{eqn:bound_reg_triang}
\end{equation}
It is worth noting that the secondary polytope is of codimension $d+1$ \cite{DeLoera2010}, whereas here we are counting the number of lattice points in a simplex of codimension one. Restricting the counting to a lower-dimensional space would yield a better upper bound, but we leave this to future work.

By restricting to FRSTs of reflexive polytopes we can apply further constraints that improve the above bound.  For this case, the relevant $n$ is the number of points not strictly interior to facets, plus the origin. The star condition requires that the component of the GKZ vector corresponding to the origin is equal to $V$, while the fine condition simply requires that all other components are positive. We have thus shown that bounding the number of FRSTs reduces to counting the number of solutions to
\begin{equation}
	\sum_{i=1}^{n-1}x_i=dV,\ \mathrm{with\ }x_i\in\ZZ_{>0}\,.
\end{equation}
Solving this simple combinatorial problem we find the bound
\begin{equation}\label{nfrst}
N_\mathrm{FRST}\leq\binom{dV-1}{n-2}\,.
\end{equation}

\begin{figure}[!ht]
    \begin{subfigure}{.5\textwidth}
    	\centering
    	\includegraphics[width=\linewidth]{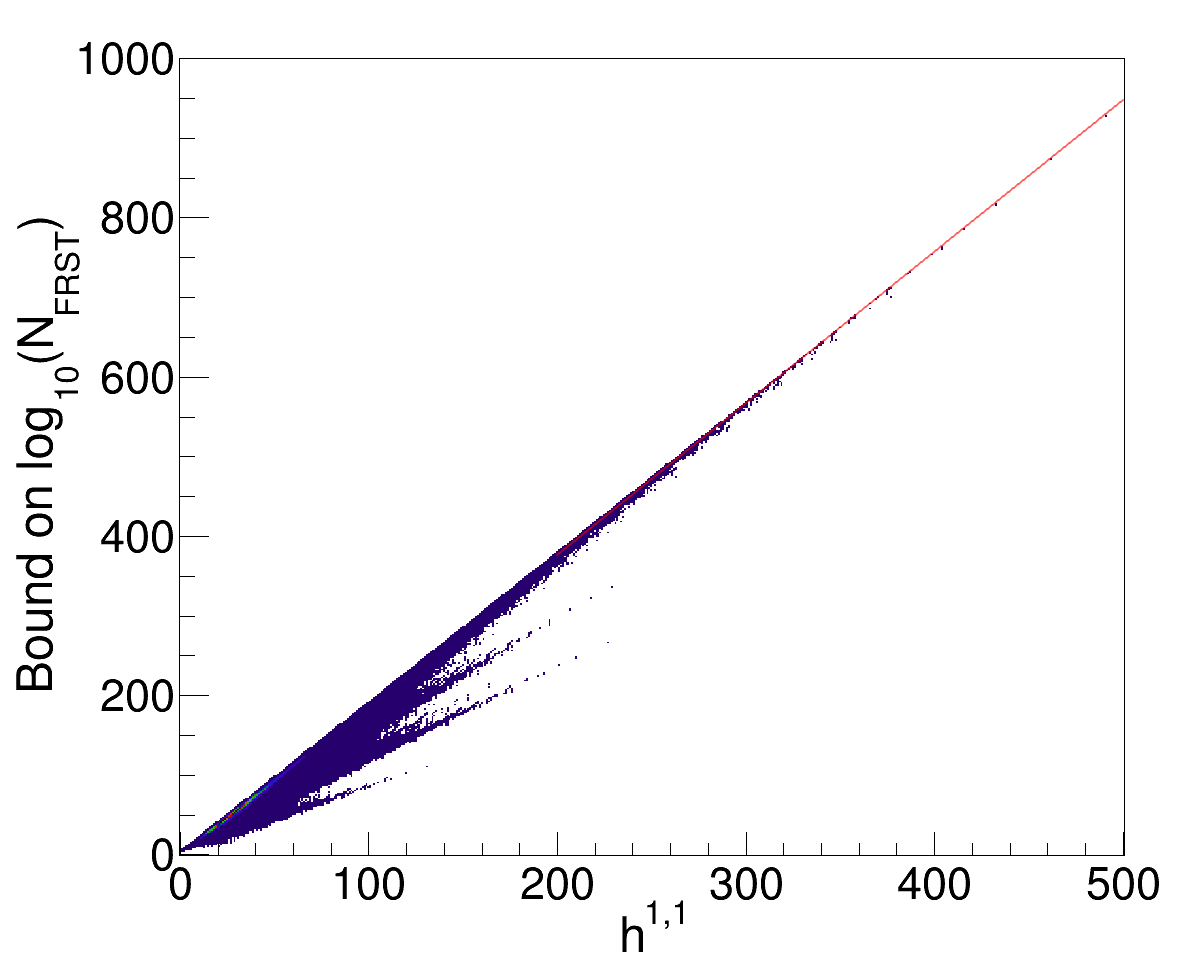}
    	\label{fig:frstbounds}
    \end{subfigure}
\begin{subfigure}{.5\textwidth}
	\centering
	\includegraphics[width=\linewidth]{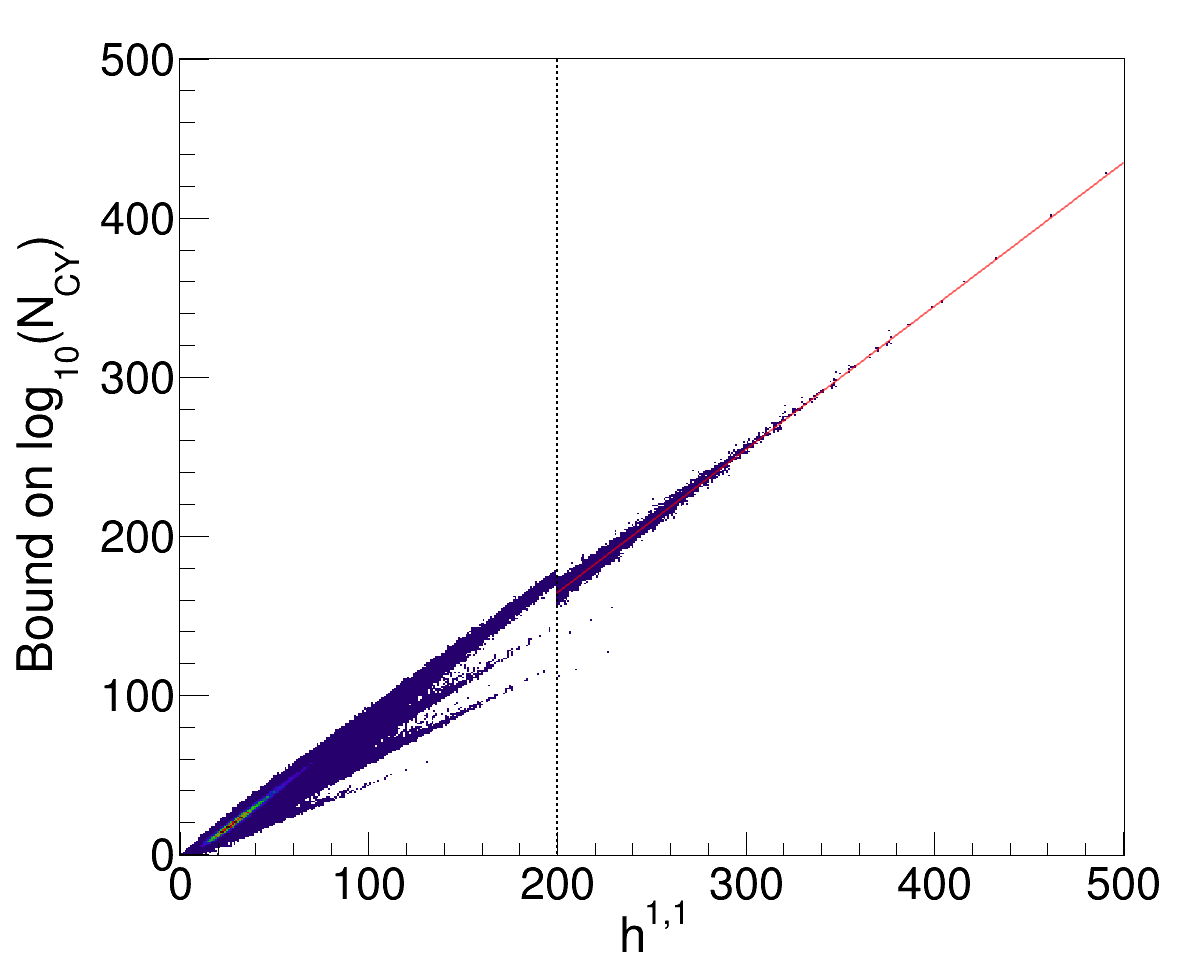}
	\label{fig:cybounds}
\end{subfigure}
	\caption{The upper bounds on the number of FRSTs, $N_{{\mathrm{FRST}}}$, and distinct Calabi-Yau threefolds, $N_{{\mathrm{CY}}}$, from four-dimensional reflexive polytopes. The fits for $h^{1,1}\geq200$, shown in red, are $\log_{10}(N_{\mathrm{FRST}}) = 1.91\,h^{1,1} -5.31$ and $\log_{10}(N_{\mathrm{CY}}) = 0.90\,h^{1,1} -15.45$.  In the right panel, we show \eqref{ncybd} for $h^{1,1}<200$, i.e.~to the left of the vertical line, while for $h^{1,1}\ge 200$ we show the tighter bound obtained by explicitly computing $N_{\mathrm{FRT}}(f)$ for some of the 2-faces $f$, and using this data in \eqref{ncybd}.}
	\label{fig:bounds_240}
\end{figure}

We can now apply the formula \eqref{nfrst} to the polytopes in the Kreuzer-Skarke database, for which $d=4$ and $n \le 496$. The resulting upper bound is shown in Figure~\ref{fig:bounds_240}, and Table~\ref{tab:bounds} shows the values for the largest polytopes. We find that the single largest polytope, which has $h^{1,1}=491$, $h^{2,1}=11$, $n=496$, and $V=3528$, gives the overwhelmingly dominant contribution to the bound: $N_{{\mathrm{FRST}}}^{(491,11)} < 1.53 \times 10^{928}$, which is 55 orders of magnitude larger than the contribution of the first subleading polytope.  This observation is consistent with \cite{Altman:2018zlc}, although our upper bound is lower than their estimate.

\subsection{Bounding the number of distinct Calabi-Yau hypersurfaces}

We have now bounded the number of FRSTs, but different FRSTs can give rise to the same Calabi-Yau threefold, in different regions of its K\"ahler moduli space.
We will now derive a stronger bound for the number of distinct --- i.e., homotopy inequivalent --- Calabi-Yau threefolds.

Wall's theorem \cite{Wall1966} implies that the homotopy types of compact Calabi-Yau threefolds with torsion-free homology are classified by the Hodge numbers, triple intersection numbers ($\kappa_{ABC}$, defined in \eqref{eq:intnums}) and second Chern class.  Now suppose that $\mathcal{T}_1$ and $\mathcal{T}_2$ are two FRSTs of a reflexive polytope $\Delta^{\circ}$ whose restrictions to the 2-faces of $\Delta^{\circ}$ are identical.  Let us show that $\mathcal{T}_1$ and $\mathcal{T}_2$
define Calabi-Yau hypersurfaces $X_1$ and $X_2$, respectively, that are homotopy equivalent.

First of all, the Hodge numbers of $X_1$ and $X_2$, being determined only by $\Delta^{\circ}$, are identical.  Similarly, the triple intersection numbers of $X_1$ and $X_2$, which are specified by the restrictions of $\mathcal{T}_1$ and $\mathcal{T}_2$ to the 2-faces of $\Delta^{\circ}$, are equal --- see for example \cite{Braun:2017nhi}.

Finally, we show that the second Chern classes of $X_1$ and $X_2$ are equal.
Viewing the second Chern class $c_2(X)$ of a threefold $X$ as a linear form acting on the homology classes of divisors $D$, $c_2[D] := \int_D c_2(X)$,
one easily shows (see e.g.~\cite{Hubsch:1992nu}) that $c_2[D]$ can be computed from the triple intersection numbers $\kappa_{ABC}$ of $X$ and the holomorphic Euler characteristic of $D$, $\chi^{h}(D):= h^{0,0}(D)-h^{0,1}(D)+h^{0,2}(D)$.
For $D$ a prime toric divisor, $\chi^{h}(D)$ is easily expressed in terms of polytope data, and so the action of $c_2(X)$ on a basis of prime toric divisors can be computed from polytope and triangulation data that is sufficient to compute the $\kappa_{ABC}$: that is, from the data of a triangulation restricted to the 2-faces of $\Delta^{\circ}$.

In summary, for a Calabi-Yau threefold hypersurface $X$ in a toric variety defined by an FRST of a reflexive polytope $\Delta^{\circ}$, the data required by Wall's theorem to determine the homotopy type of $X$ is specified by the \emph{restriction of the FRST to the 2-faces} of $\Delta^{\circ}$.  In particular, two FRSTs of $\Delta^{\circ}$ that differ only by simplices in 3-faces define homotopy equivalent Calabi-Yau threefolds.

Thus, to bound the number of homotopy inequivalent Calabi-Yau hypersurfaces arising from the Kreuzer-Skarke list, we need only bound the number of triangulations of the 2-faces of the corresponding polytopes.  This reduces our task to that of counting triangulations of lattice polygons, which is rather well-understood.  To our knowledge, the following theorem by Anclin \cite{Anclin2003} provides the best upper bound for our purpose. We use the more convenient formulation presented in Theorem 9.3.7 of  \cite{DeLoera2010}.

\begin{theorem}
    \emph{(Anclin 2003)} The number of fine triangulations of a lattice polygon is bounded by $2^{3i+b-3}$, where $i$ and $b$ are the number of interior and boundary lattice points in the polygon.\footnote{This theorem is applicable to lattice polygons embedded in a higher-dimensional lattice since their affine $\mathbb{Z}$-span is an affine lattice isomorphic to $\mathbb{Z}^2$.}
\end{theorem}
We stress that this is an upper bound for the number of all fine, not necessarily regular, triangulations.  Even so, it yields a stronger bound than the method described in \S\ref{sec:bound_lat_poly}, even when the latter is restricted to fine and regular triangulations.

From the above theorem it follows that the number of distinct Calabi-Yau threefolds that arise from a particular four-dimensional reflexive polytope is bounded by
\begin{equation}\label{ncybd}
    N_\mathrm{CY}\leq \prod_{f\in F_{\Delta^\circ}(2)}2^{3i_f+b_f-3},
\end{equation}
where $F_{\Delta^\circ}(2)$ denotes the set of 2-faces of $\Delta^\circ$ and $i_f$ and $b_f$ denote the number of interior and boundary points, respectively, of the 2-face $f$.   Sometimes one can compute the actual number $N_{\mathrm{FRT}}(f)$ of fine, regular triangulations for certain 2-faces $f$, such as those with relatively few points, and replace the corresponding factors in \eqref{ncybd} with $N_{\mathrm{FRT}}(f)$, leading to a tighter bound.

We have evaluated \eqref{ncybd} for the 473{,}800{,}776 polytopes in the Kreuzer-Skarke list.  For polytopes with $h^{1,1}\geq 200$, we used {\tt TOPCOM} \cite{Rambau2002} to compute $N_{\mathrm{FRT}}(f)$ for 2-faces $f$ with up to 17 lattice points,\footnote{We stop at 17 lattice points because in the polytope with $h^{1,1}=491$, which dominates the bound, the next-larger 2-face has 129 lattice points, making an exact count impractical.} while for the far more numerous polytopes with small $h^{1,1}$ we used \eqref{ncybd} directly.

In total, we find that the number of distinct Calabi-Yau threefolds arising from the Kreuzer-Skarke list obeys\footnote{The exact bound is 16468748345409819248194378023304313940946564187432078509364852963014
6040575488255243589332232903592518474658158429571147325462793749362868946853319691948129
8883395339972203613804284096289742550789265088421675080072368874768688779979297016169174
5194720127658754222186509852391225057750610021878875802809122461761270655713263315386424
8913905073354395569564592722506413231664691489296656747662315313727961454833931913041383
322595623.}
\begin{equation}\label{thebd}
N_\mathrm{CY} < 1.65 \times 10^{428}\,.
\end{equation}
This is one of our main results.

The bound \eqref{thebd} is dominated by the polytope with $h^{1,1}=491$, which contains the largest 2-face in the database, and contributes
\begin{equation}\label{the491bd}
N_\mathrm{CY}^{(491,11)} \le \Bigl(2^{6} \cdot 3^2 \cdot 5^2 \cdot 17^2 \cdot 97^2 \cdot 101^2\Bigr)\cdot 2^{1374} \approx 1.65 \times 10^{428}\,,
\end{equation} where the factors in parentheses come from the numbers of triangulations of 2-faces with $\le 17$ points.
The contribution \eqref{the491bd} is 27 orders of magnitude more than that from the first subleading polytope.

In Figure~\ref{fig:bounds_240} we show the contributions to the upper bound as a function of $h^{1,1}$, and explicit values for the largest polytopes are shown in Table~\ref{tab:bounds}.

\begin{table}[ht]
\centering
	\caption{\label{tab:bounds}Upper bounds for polytopes with $h^{1,1}\geq 400$.}
		\begin{tabularx}{0.7\textwidth}{*{4}{Y}}
			\toprule
			\multicolumn{2}{c}{Hodge Numbers}
			& \multicolumn{2}{c}{Upper bounds}\\
			\cmidrule(lr){1-2} \cmidrule(l){3-4}
			$h^{1,1}$ & $h^{2,1}$ & $N_\text{FRST}$ & $N_\text{CY}$\\
			\midrule
			491 & 11 & $1.53\times 10^{928}$ & $1.65\times 10^{428}$\\
			\midrule
			\multirow{2}{*}{462} &  \multirow{2}{*}{12}
			&   $1.13\times 10^{873}$ & $3.71\times 10^{401}$\\
			& & $2.67\times 10^{873}$ & $2.66\times 10^{401}$\\
			\midrule
			\multirow{4}{*}{433} & \multirow{4}{*}{13}
			&   $8.39\times 10^{817}$ & $8.36\times 10^{374}$\\
			& & $1.98\times 10^{818}$ & $6.00\times 10^{374}$\\
			& & $4.67\times 10^{818}$ & $4.30\times 10^{374}$\\
			& & $3.06\times 10^{816}$ & $3.00\times 10^{374}$\\
			\midrule
			\multirow{2}{*}{416} & \multirow{2}{*}{14}
			&   $2.46\times 10^{786}$ & $7.64\times 10^{359}$\\
			& & $1.04\times 10^{786}$ & $6.65\times 10^{359}$\\
			\midrule
			\multirow{6}{*}{404} & \multirow{6}{*}{14}
			&   $1.47\times 10^{763}$ & $1.35\times 10^{348}$\\
			& & $3.47\times 10^{763}$ & $9.69\times 10^{347}$\\
			& & $8.20\times 10^{763}$ & $6.94\times 10^{347}$\\
			& & $2.27\times 10^{761}$ & $6.75\times 10^{347}$\\
			& & $5.37\times 10^{761}$ & $4.84\times 10^{347}$\\
			& & $8.30\times 10^{759}$ & $2.42\times 10^{347}$\\
			\bottomrule
	\end{tabularx}
\end{table}

As a final comment, even though there exist lower bounds for the number of triangulations of polygons, e.g.~in \cite{Kaibel03}, these do not directly lead to lower bounds on the number of Calabi-Yau threefolds, as not all combinations of triangulations of 2-faces occur in a full triangulation of the polytope.

\section{Random Sampling of Calabi-Yau Threefolds}\label{fullsec:sampling}

The set of Calabi-Yau threefold hypersurfaces in toric varieties is potentially of astronomical proportions.  As it is not possible to study every Calabi-Yau hypersurface in this set, we instead aim to construct representative ensembles of hypersurfaces. Below we describe a method that allows us to produce diverse samples, and we demonstrate that it in fact produces \emph{fair} samples, at least at the modest values of $h^{1,1}$ where we can test it.
The key strategy that makes this possible is to explore the space of regular triangulations via their height vectors, as well as with the more common technique of performing bistellar flips.

In this section, we first describe our sampling algorithms and the software used to implement them. We then use these algorithms to obtain random samples of Calabi-Yau threefolds arising from the polytopes with largest Hodge numbers. Finally, we analyze these ensembles and present the statistics of relevant geometric quantities.

\subsection{Sampling algorithms}\label{sec:sampling}

The most straightforward way of obtaining random triangulations is to move in the space of triangulations by performing random bistellar flips (see e.g. \cite{caputo2015,Kaibel03}).

\vspace{0.3cm}
\begin{algorithm}[H]
    \begin{enumerate}
        \item Pick an initial FRST $\cT = \cT_\circ$.
        \item Construct a new FRST $\cT_\text{new}$ by performing $N_\text{flip}$ random flips on $\cT$, making sure that the resulting triangulation is fine, regular and star.
        \item Add $\cT_\text{new}$ to the sample, set $\cT = \cT_\text{new}$ and repeat step 2.
    \end{enumerate}
    \caption{Random Flips}
    \label{algo:random_1}
\end{algorithm}
\vspace{0.3cm}

The main shortcoming of this algorithm is that it has a very long mixing time: it produces samples that are strongly dependent on the initial triangulation $\cT_\circ$, and the statistics of the sample converge very slowly, even for relatively small polytopes with $\cO(10)$ points. The reason for this behavior is that Algorithm \ref{algo:random_1} explores the space of FRSTs by taking small steps, performing $N_\text{flip}$ flips at a time, and it takes a long time to move from one corner of the space to the other. Increasing the step size by setting $N_\text{flip} \gg 1$ is prohibitively expensive, as regularity needs to be checked after each flip.

We now propose an algorithm that addresses this issue. As explained in \S\ref{sec:reg_traings_and_sec_poly}, Theorem \ref{theorem:height_paths} implies that the set of heights $\mathbf{h}$ that result in an FRST form a convex, polyhedral cone, which we denote by $\widetilde{\mathcal{C}}$. We can thus explore the space of FRSTs by taking random walks within $\widetilde{\mathcal{C}}$. Moreover, since a homogeneous scaling of the heights $\mathbf{h} \rightarrow \lambda \mathbf{h}, \, \lambda \in \mathbb{R}_+$ does not change the triangulation, a suitable subspace to explore is the intersection of $\widetilde{\mathcal{C}}$ and a sphere of radius one:  $\mathcal{C} := \{ \mathbf{h} | \mathcal{T}_h \, \text{is fine and star, and}\,|{\mathbf{h}}|=1\}$. Although constructing $\cC$ explicitly is difficult, we can obtain a random sample of points in $\cC$ using a hit and run algorithm, as it is easy to determine whether a given height vector $\mathbf{h}$ is contained in $\cC$. In practice, we explore the space of fine, regular but not necessarily star triangulations this way, and transform them into FRSTs at the end, for technical reasons we describe in \S\ref{sec:software}.

\vspace{0.3cm}
\begin{algorithm}[H]
    \begin{enumerate}
        \item Pick an initial height vector $\mathbf{h}=\mathbf{h}_\circ \in \mathcal{C}$.
        \item Pick a random unit vector $\boldsymbol{\alpha} \in \mathbb{R}^n$, $|\boldsymbol{\alpha}|=1$.
        \item Consider $\mathbf{h}' = (\mathbf{h} + \epsilon \boldsymbol{\alpha}) / |\mathbf{h} + \epsilon \boldsymbol{\alpha}|$, and find the largest $\epsilon = \epsilon_\text{max}$ such that $\mathbf{h}' \in \mathcal{C}$.
        \item Pick a random real number $\delta \in [0, \epsilon_\text{max}]$ and set $\mathbf{h} =(\mathbf{h} + \delta \boldsymbol{\alpha}) / |\mathbf{h} + \delta \boldsymbol{\alpha}|$.
        \item Repeat steps 2-4, $N_\text{walk}$ times.
        \item Compute $\cT_{h}$, convert it to a star triangulation, and add it to the sample.
        \item Repeat steps 2-6.
    \end{enumerate}
    \caption{Random Walks}
    \label{algo:random_2}
\end{algorithm}
\vspace{0.3cm}

The main advantage of Algorithm \ref{algo:random_2} is that it converges much faster than Algorithm \ref{algo:random_1}. There are two factors that contribute to this improvement. First, obtaining a regular triangulation from a given height vector is significantly faster than determining whether a given triangulation is regular. Second, a single step of Algorithm \ref{algo:random_2} can produce FRSTs that differ by a large number of bistellar flips, enabling more efficient exploration of the space of FRSTs.

A crucial step in implementing Algorithm \ref{algo:random_2} is finding the initial height vector $\mathbf{h}_\circ \in \mathcal{C}$. A natural choice is $\mathbf{h}_\circ^i = |\mathbf{p}^i|^2$ where $|\mathbf{p}^i|$ are the norms of the lattice points on the polytope. The height vector $\mathbf{h}_\circ$ is not generic and gives rise to a subdivision, rather than a triangulation. A triangulation can be obtained by an infinitesimal perturbation $\mathbf{h} = \mathbf{h}_\circ + \epsilon$. Such triangulations are called Delaunay, and they enjoy a number of nice properties, including being fine and regular \cite{DeLoera2010}.

Although Algorithm \ref{algo:random_2} is very efficient, it has a crucial shortcoming: even if it returns a representative sample of points in $\mathcal{C}$, the resulting sample of FRSTs is not  representative. In fact, different regions in $\mathcal{C}$ corresponding to different triangulations need not have the same volume, cf.~Figure \ref{fig:491_xsec}, and we expect the triangulations corresponding to larger regions to be sampled more frequently. One way to combat this problem is by estimating the volumes of these regions and constructing a weighted sample of triangulations. However, this becomes prohibitively expensive at large $h^{1,1}$. We will instead address this issue by combining Algorithms \ref{algo:random_1} and \ref{algo:random_2}:

\vspace{0.3cm}
\begin{algorithm}[H]
    \begin{enumerate}
        \item Pick an initial height vector $\mathbf{h}=\mathbf{h}_\circ \in \mathcal{C}$.
        \item Pick a random vector unit vector $\boldsymbol{\alpha} \in \mathbb{R}^n$, $|\boldsymbol{\alpha}|=1$.
        \item Consider $\mathbf{h}' = (\mathbf{h} + \epsilon \boldsymbol{\alpha}) / |\mathbf{h} + \epsilon \boldsymbol{\alpha}|$, and find the largest $\epsilon = \epsilon_\text{max}$ such that $\mathbf{h}' \in \mathcal{C}$.
        \item Pick a random real number $\delta \in [0, \epsilon_\text{max}]$ and set $\mathbf{h} =(\mathbf{h} + \delta \boldsymbol{\alpha}) / |\mathbf{h} + \delta \boldsymbol{\alpha}|$.
        \item Repeat steps 2-4, $N_\text{walk}$ times.
        \item Compute $\cT_{h}$.
        \item Construct a new triangulation $\cT_{\text{new}}$ by performing $N_\text{flips}$ random flips on $\cT_{h}$, making sure that the triangulation stays fine and regular.
        \item Convert $\cT_\text{new}$ to a star triangulation, and add it to the sample.
        \item Repeat steps 2-8.
    \end{enumerate}
    \caption{Random Walks + Random Flips}
    \label{algo:random_3}
\end{algorithm}
\vspace{0.3cm}
This algorithm does not suffer from the weaknesses mentioned above.  As we explain in Appendix \ref{sec:app_randomsample}, at small $h^{1,1}$ we can enumerate all the distinct Calabi-Yau hypersurfaces resulting from a polytope, and verify that the output of Algorithm \ref{algo:random_3} gives a fair sampling of this set.  We cannot similarly prove that Algorithm \ref{algo:random_3} gives a fair sampling for polytopes with $h^{1,1} \gg 1$, because computing the true distribution for the purpose of verification is too expensive.  However, we are not aware of any particular biases in the sampling, and so we propose Algorithm \ref{algo:random_3} as a way to efficiently generate (provisionally) fair samples for any polytope in the Kreuzer-Skarke list.

\subsection{Geometric data} \label{sec:geometric_data}
Before constructing random samples of Calabi-Yau threefolds using the algorithms described above, we recall the definitions of the relevant geometric data of a  Calabi-Yau hypersurface in a toric variety.

Let $\Delta^\circ$ be a four-dimensional reflexive polytope, and consider the set of lattice points $\{v^A\} \in \mathbb{Z}^4$ that are on the boundary of $\Delta^\circ$ but are not strictly interior to a facet. An FRST of $\{v^A\}$ defines a toric variety $V$ of which a generic anticanonical hypersurface $X \subset V$ is a smooth Calabi Yau threefold. The Hodge numbers $h^{1,1}$ and $h^{2,1}$ of $X$ are determined solely by polytope data. In particular, when $\Delta^\circ$ is favorable,\footnote{See e.g.~\S2.1.1 of \cite{Demirtas:2018akl} for the definition.}
which is the case for all polytopes we consider below, there are $h^{1,1}+4$ lattice points $\{v^A\}$.

Each of the points $v^A$ corresponds to a prime effective divisor $\widehat{D}_A$ of $V$, and the intersection $D_A = \widehat{D}_A \cap X$ defines an irreducible effective divisor on $X$.\footnote{In general, not all irreducible effective divisors of $X$ can be written this way: see e.g.~\cite{Demirtas:2018akl}.} The irreducible algebraic curves $C_a$ on $X$ generate the Mori cone, $\cM_X \subset H_2(X,\mathbb{Z})$.  The volumes of algebraic subvarieties of $X$ are calibrated by the K\"ahler form $J$,
\begin{equation}
    \begin{aligned}
        \cV := \text{Vol}(X, J) &= \frac{1}{6}\int_X J \wedge J \wedge J\, , \\
        \tau_A := \text{Vol}(D_A, J) &= \frac{1}{2} \int_{D_A} J \wedge J\, , \\
        \mathfrak{t}_a:=\text{Vol}(C_a, J) &= \int_{C_a} J\, .
    \end{aligned}
    \label{eq:volumes1}
\end{equation}
The K\"ahler cone of $X$, $\cK_X \subset H^2(X,\mathbb{R})$, is defined as the set of cohomology classes of closed $(1,1)$ forms $J$ for which the volumes above are all positive. We define the intersection numbers
\begin{equation}
    \begin{aligned}
        M_{aA} &= \# C_a \cap D_A\, , \\
        \kappa_{ABC} &= \# D_A \cap D_B \cap D_C \, .
    \end{aligned}
\label{eq:intnums}
\end{equation}
It is useful to pick $h^{1,1}$ of the irreducible divisors $D_i, i=1,\dots,h^{1,1}$, to construct a basis for $H_4(X,\mathbb{Z})$. The Poincare duals $[D_i]$ of the $D_i$ furnish a complete basis for $H^2(X,\mathbb{Z})$. We expand the K\"ahler form $J$ in terms of $[D_i]$,
\begin{equation}
    J = t^i [D_i]\, ,
\end{equation}
and rewrite \eqref{eq:volumes1} as
\begin{equation}
    \begin{aligned}
        \cV &= \frac{1}{6} \kappa_{ijk} t^i t^j t^k \, , \\
        \tau_A &= \frac{1}{2} \kappa_{Ajk} t^j t^k  \, , \\
        \mathfrak{t}_a &= M_{ak} t^k \, .
    \end{aligned}
    \label{eq:volumes2}
\end{equation}
The K\"ahler metric $K_{ij}$ computed from the K\"ahler potential $\cK = -2 \log{\cV}$ can be expressed as
\begin{equation}
    K_{ij} = \frac{t_i t_j}{8 \cV^2 } - \frac{A_{ij}}{4\cV}\,,
    \label{eq:Kahler}
\end{equation}
where $A_{ij} = \frac{\partial t_i}{\partial \tau_j}$.

For most Calabi-Yau hypersurfaces $X$, computing the generators of the Mori cone $\cM_X$ is difficult. The Mori cone of the ambient variety $\cM_V \supset \cM_X$, however, is easily computed from toric data and is a useful approximation for $\cM_X$. As different ambient toric varieties can give rise to the same Calabi-Yau hypersurface, a better approximation to $\cM_X$ can be obtained by taking the intersection of the Mori cones of the ambient varieties,
\begin{equation}
    \cM_\cap :=  \underset{\alpha}{\bigcap} \, \cM_{V_\alpha}\, .
\end{equation}
Here, $\alpha$ runs over the ambient toric varieties that contain the same Calabi-Yau hypersurface. This is equivalent to gluing together the K\"ahler cones of the ambient varieties,
\begin{equation}
    \cK_\cup :=  \underset{\alpha}{\bigcup} \, \cK_{V_\alpha}\, .
    \label{eq:kcupcone}
\end{equation}

To study the large-volume region of the K\"ahler cone $\cK_V$, we define the \emph{stretched K\"ahler cone} $\widetilde{\cK_V}$ as the subset of $\cK_V$ where the volumes of all curves $C_a \in \cM_V$ are greater than one,
\begin{equation}
    \widetilde{\cK_V} = \Bigl\{ J \in H^{1,1}(X,\mathbb{R}) \,\Bigl|\, \mathfrak{t}_a \geq 1~~\forall~C_a \in \cM_V \Bigr. \Bigr\}\, .
    \label{eq:strkcone}
\end{equation}
Similarly, we define
\begin{equation}
    \widetilde{\cK_\cup} = \Bigl\{ J \in H^{1,1}(X,\mathbb{R}) \,\Bigl|\, \mathfrak{t}_a \geq 1~~\forall~C_a \in \cM_\cap \Bigr. \Bigr\}\, .
    \label{eq:strkcupcone}
\end{equation}
In practice, constructing $\widetilde{\cK_\cup}$ is prohibitively expensive at large $h^{1,1}$. Thus, we will construct and use $\widetilde{\cK_V}$ in most of our analyses.

Finally, we define \textit{the tip of the stretched K\"ahler cone} as the point $J_\star \in \widetilde{\cK_V}$ that is closest to the origin, and we write
\begin{equation}
   J_\star = t_\star^i [D_i]\,.
   \label{eq:strkconetip}
\end{equation}
The distance between the origin and $J_\star$ is then
\begin{equation}
    d_\text{min} = \sqrt{\delta_{ij} t_\star^i t_\star^j}.
    \label{eq:dmin}
\end{equation}
Many physical quantities of interest depend on the K\"ahler form $J$.  Ideally, given a mechanism for moduli stabilization at large $h^{1,1}$ one might sample K\"ahler forms within the K\"ahler cone according to a measure defined by the dynamics of stabilization, i.e.~favoring regions of moduli space where stabilization is likely.  Here, for want of a sufficiently general theory of moduli stabilization, we will set $J \to J_\star$.  Thus, we will compute $\cV$ and the eigenvalues of $K_{ij}$ at the tip of the stretched K\"ahler cone.  Revisiting our analysis in the context of specific scenarios for moduli stabilization would be worthwhile.

\subsection{Computation}\label{sec:software}

Obtaining random samples of Calabi-Yau hypersurfaces and computing their topological data poses a significant computational challenge, especially at large $h^{1,1}$. To this end, we have developed a number of algorithms and made use of a collection of software packages, which we now summarize.

Given the vertices of a reflexive polytope, we first find the lattice points not strictly interior to facets using {\tt PALP} \cite{Kreuzer_2004}. (An alternative library for analyzing polytopes is {\tt PPL} \cite{BAGNARA20083}. For a convenient wrapper, see {\tt SageMath} \cite{sagemath}.) We then obtain a fine and regular triangulation by lifting these points into one extra dimension by a set of heights, constructing the convex hull using {\tt Qhull} \cite{Barber96}, and projecting down its lower faces to obtain the simplices of the triangulation.\footnote{Within the space of height vectors, there will be lower-dimensional subspaces that give rise to convex hulls with non-simplicial faces, which {\tt Qhull} automatically subdivides into simplices. This subdivision may not preserve regularity, so one must avoid such cases.} To obtain a star triangulation, the height of the origin can be set to be much lower than all other heights. However, for large polytopes this leads to numerical overflow issues with {\tt Qhull}. As a solution, we first construct a non-star triangulation and make it star by removing edges strictly interior to the polytope and connecting all points to the origin, as described in \cite{Braun:2014xka}.\footnote{It is worth mentioning that there is a significantly faster triangulation library in the Computer Geometry Algorithms Library (CGAL) \cite{cgal:eb-19a}. With it, one can directly compute star triangulations, even for the largest polytopes in the Kreuzer-Skarke database. Additionally, when further subdivisions are necessary for non-generic height vectors, it always performs one that preserves regularity. More details can be found in \cite{cgal:dDTriangs}.}

Although the triangulation algorithm described above is very fast, it does not allow one to systematically compute all triangulations of a given point configuration. We use {\tt TOPCOM} \cite{Rambau2002} and {\tt MPTOPCOM} \cite{Jordan2018} for this purpose. Carrying out the random sampling Algorithms \ref{algo:random_1} and \ref{algo:random_3} requires performing random flips and checking regularity. We use a modified version of {\tt TOPCOM} to perform random flips. Checking whether a triangulation is regular using these packages, however, is prohibitively slow. To perform this check, we compute the corresponding secondary cone, which is full-dimensional if and only if the triangulation is regular \cite{Gelfand1989,Gelfand1991}. The problem of determining whether a cone is full-dimensional can be translated into checking the feasibility of a linear program, which we carry out using Google's {\tt OR-Tools} \cite{ortools}.

Once we obtain an FRST, the next step is to compute the topological data of the Calabi-Yau hypersurface. We compute the generators of the Mori cone, cf.~\eqref{eq:intnums}, using the algorithm described in \cite{Berglund1995} (see also \cite{oda1991}).  Finding the tip of the stretched K\"ahler cone is a sparse quadratic programming problem, which we solve using the optimization package {\tt Mosek} \cite{mosek}. To compute the triple intersection numbers $\kappa_{ijk}$ of the hypersurface, we first compute the intersection numbers $\kappa_{ABCD}$ of the prime effective divisors in the ambient variety. When $A,B,C,D$ are all distinct, we calculate the corresponding intersection number simply by inspecting the simplices in the FRST. We then use the linear equivalences between the prime effective divisors to construct a linear system of equations involving the self-intersection numbers.\footnote{We are indebted to Mike Stillman for this method of calculating the intersection numbers.} The resulting system is overdetermined but consistent, and can be solved using {\tt Mathematica} \cite{Mathematica}. (For an alternative tool for solving sparse linear systems, see \cite{10.1145/1391989.1391995}.)

\subsection{Random samples at large Hodge numbers}

We now use the methods described above to construct and study random ensembles of Calabi-Yau hypersurfaces arising from three exceptional polytopes, with the Hodge numbers $(h^{1,1},h^{2,1})=(491,11)$, $(11,491)$, and $(251,251)$.  These are the three polytopes with the maximal value of $h^{1,1}+h^{2,1}$ in the Kreuzer-Skarke list, and the first and the second, which are dual to each other, attain the maximal values of $h^{1,1}$ and $h^{2,1}$, respectively.

These polytopes, which we denote by $\Delta^\circ_{h^{1,1},h^{2,1}}$, are defined by the vertices
\begin{align}
    \hspace{-0.12cm}\Delta^\circ_{491,11}\hspace{-0.1cm}:\hspace{-0.1cm}
    \begin{pmatrix}
        v_1 & v_2 & v_3 & v_4 & v_5\\
        \hline
        1 & 0 & 0 & 21 & -63 \\
        0 & 1 & 0 & 28 & -56 \\
        0 & 0 & 1 & 36 & -48 \\
        0 & 0 & 0 & 42 & -42 \\
    \end{pmatrix}
    \hspace{0.3cm} \Delta^\circ_{251,251}\hspace{-0.1cm}:\hspace{-0.1cm}
    \begin{pmatrix}
        v_1 & v_2 & v_3 & v_4 & v_5\\
        \hline
        1 & 0 & 0 & 0 & -903 \\
        0 & 1 & 0 & 0 & -602 \\
        0 & 0 & 1 & 0 & -258 \\
        0 & 0 & 0 & 1 & -42 \\
    \end{pmatrix}
    \hspace{0.3cm}  \Delta^\circ_{11,491}\hspace{-0.1cm}:\hspace{-0.1cm}
    \begin{pmatrix}
        v_1 & v_2 & v_3 & v_4 & v_5\\
        \hline
        1 & 0 & 0 & 0 & -42 \\
        0 & 1 & 0 & 0 & -28 \\
        0 & 0 & 1 & 0 & -12 \\
        0 & 0 & 0 & 1 & -1 \\
    \end{pmatrix}
\end{align}
All three polytopes are favorable, and $\Delta^\circ_{251,251}$ is self-dual.

\begin{figure}
\begin{subfigure}{.5\textwidth}
  \centering
  \includegraphics[width=.9\linewidth]{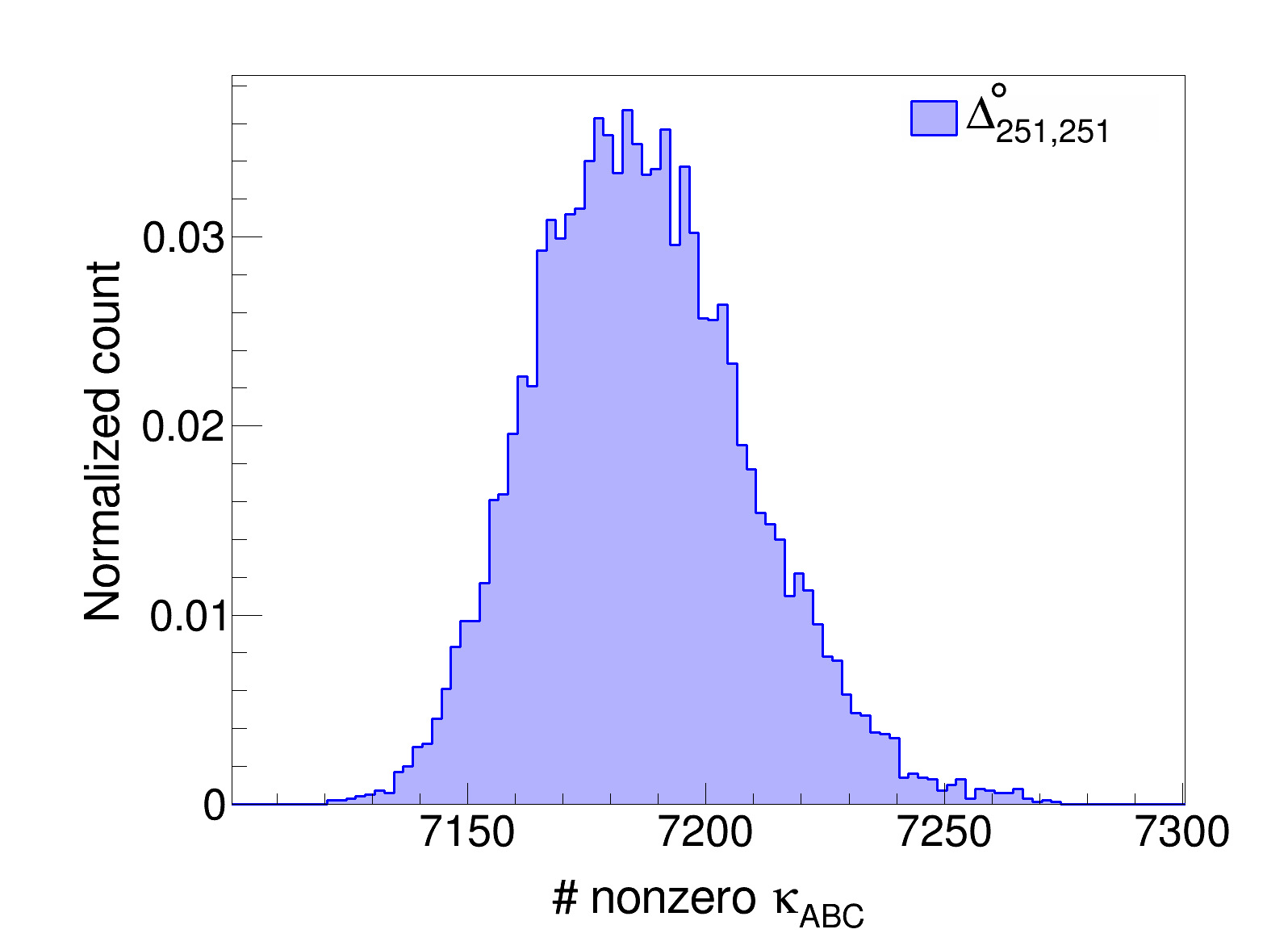}
  \caption{$\#$ of nonzero entries of $\kappa_{ABC}$}
  \label{fig:intnumnonzero251}
\end{subfigure}
\begin{subfigure}{.5\textwidth}
  \centering
  \includegraphics[width=.9\linewidth]{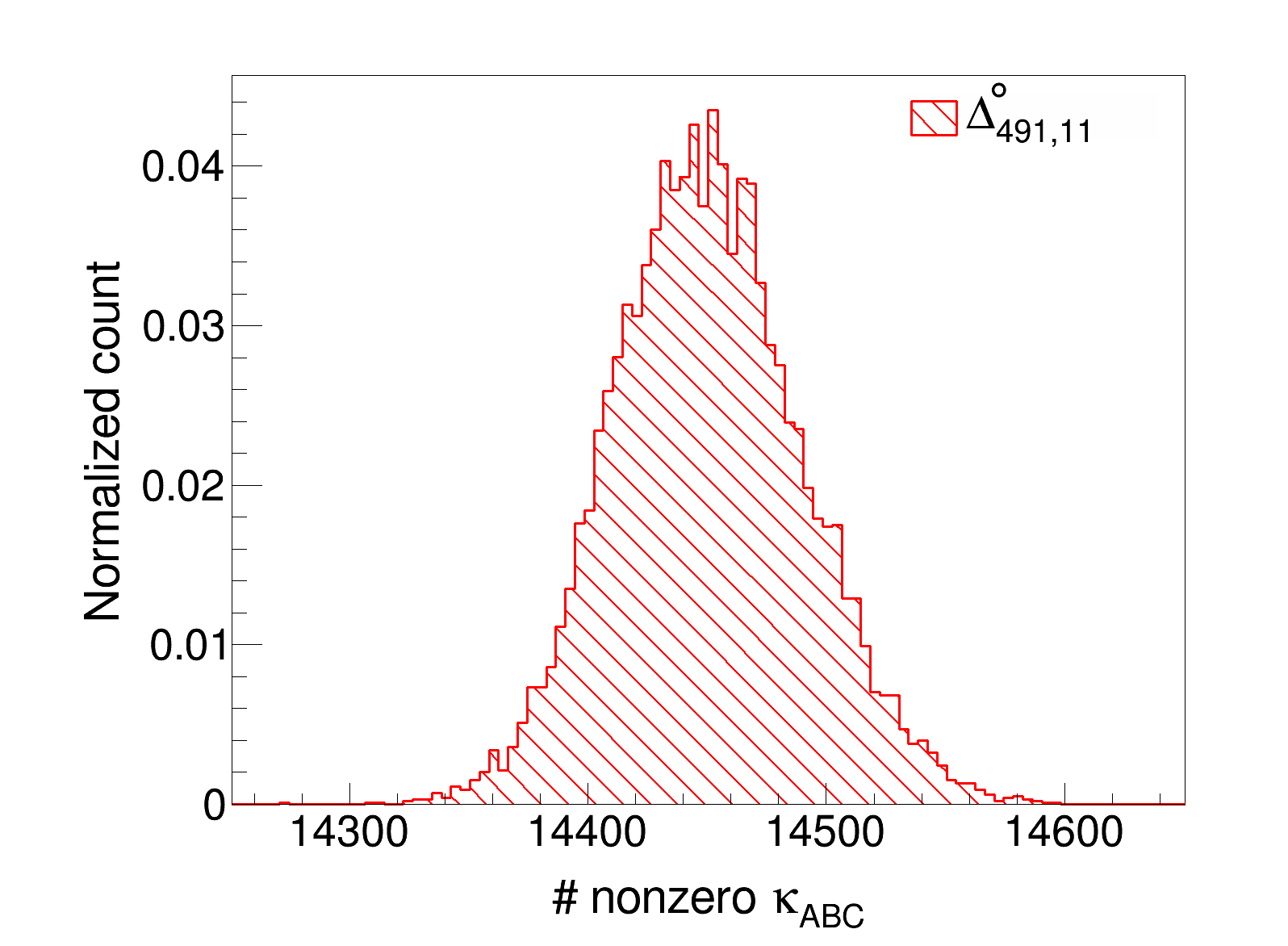}
  \caption{$\#$ of nonzero entries of $\kappa_{ABC}$}
  \label{fig:intnumnonzero491}
\end{subfigure}
\begin{subfigure}{.5\textwidth}
  \centering
  \includegraphics[width=.9\linewidth]{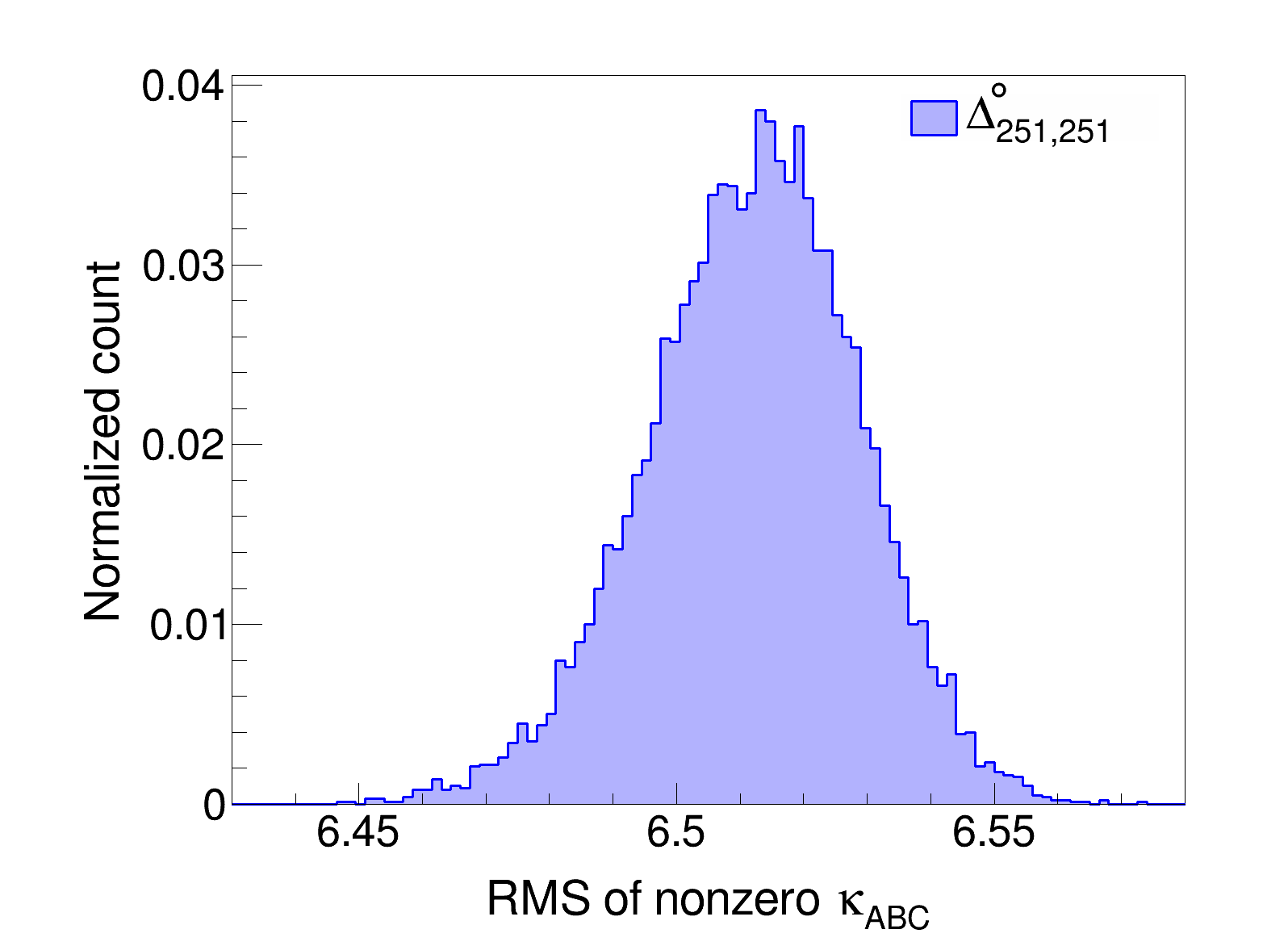}
  \caption{RMS size of nonzero entries of $\kappa_{ABC}$}
  \label{fig:intnumrms251}
\end{subfigure}
\begin{subfigure}{.5\textwidth}
  \centering
  \includegraphics[width=.9\linewidth]{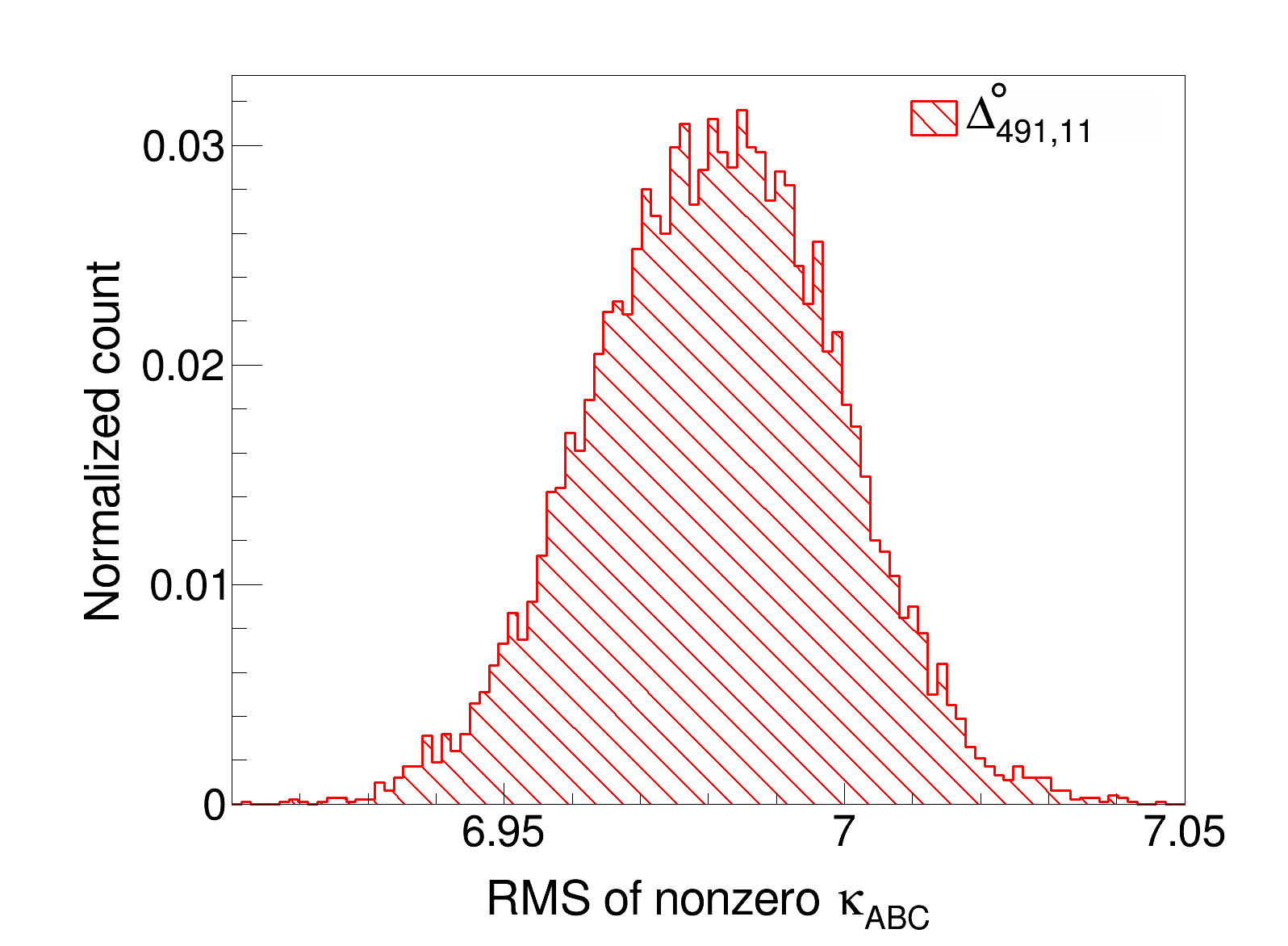}
  \caption{RMS size of nonzero entries of $\kappa_{ABC}$}
  \label{fig:intnumrms491}
\end{subfigure}
\caption{Distributions of the number and RMS size of the nonvanishing intersection numbers $\kappa_{ABC}$, cf. \eqref{eq:intnums}, from a sample of 10,000 triangulations of $\Delta^\circ_{491,11}$ and $\Delta^\circ_{251,251}$, obtained using Algorithm~\ref{algo:random_3} with $N_\text{flip}=40$ and $N_\text{walk}=50$.}
\label{fig:intnum}
\end{figure}

\begin{figure}
\begin{subfigure}{.5\textwidth}
  \centering
  \includegraphics[width=\linewidth]{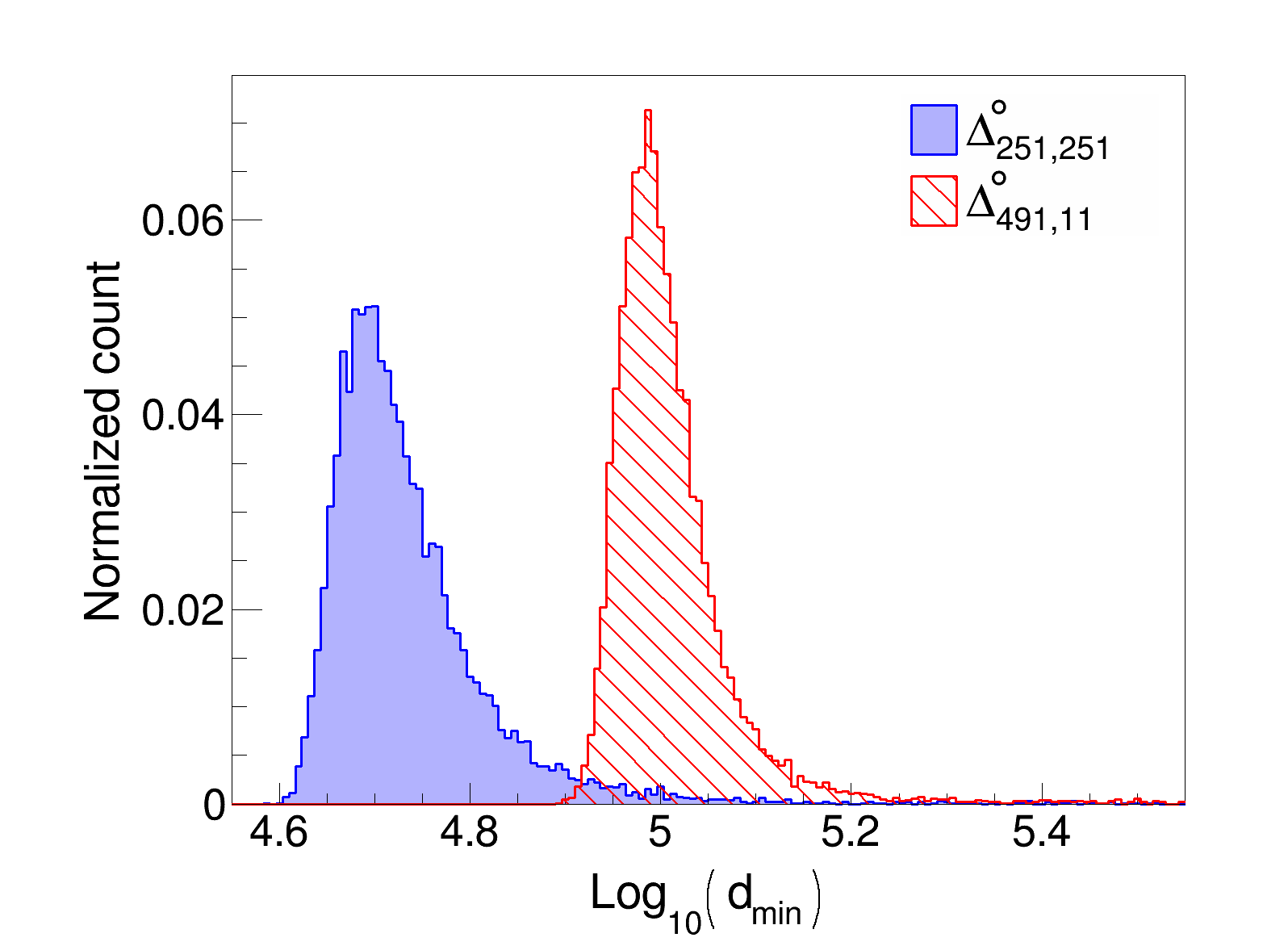}
  \caption{$d_\mathrm{min}$}
  \label{fig:dmin_251-491}
\end{subfigure}
\begin{subfigure}{.5\textwidth}
  \centering
  \includegraphics[width=\linewidth]{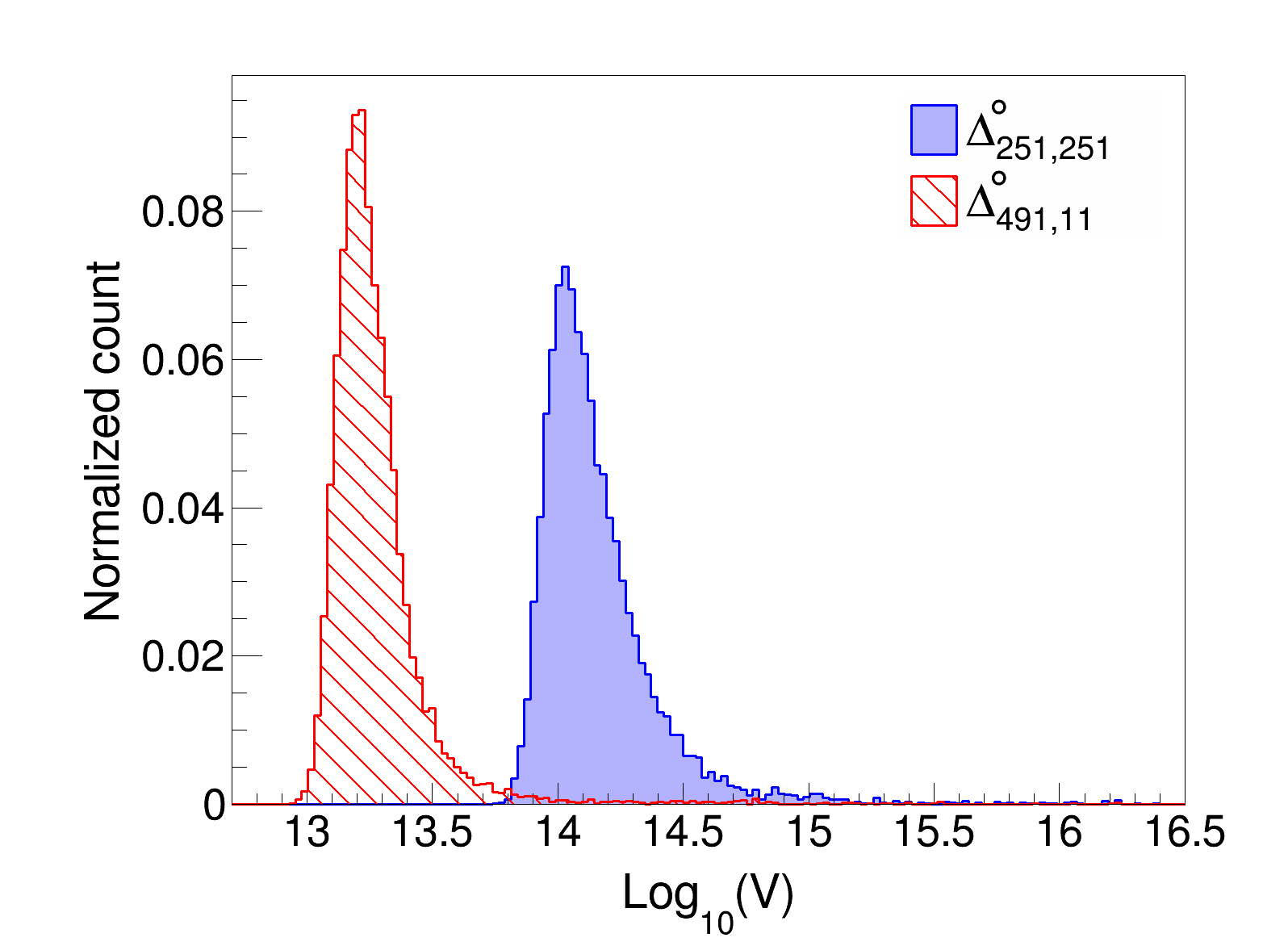}
  \caption{$\cV_{J_\star}$}
  \label{fig:cyvol_251-491}
\end{subfigure}
\par\bigskip
\begin{subfigure}{.5\textwidth}
  \centering
  \includegraphics[width=\linewidth]{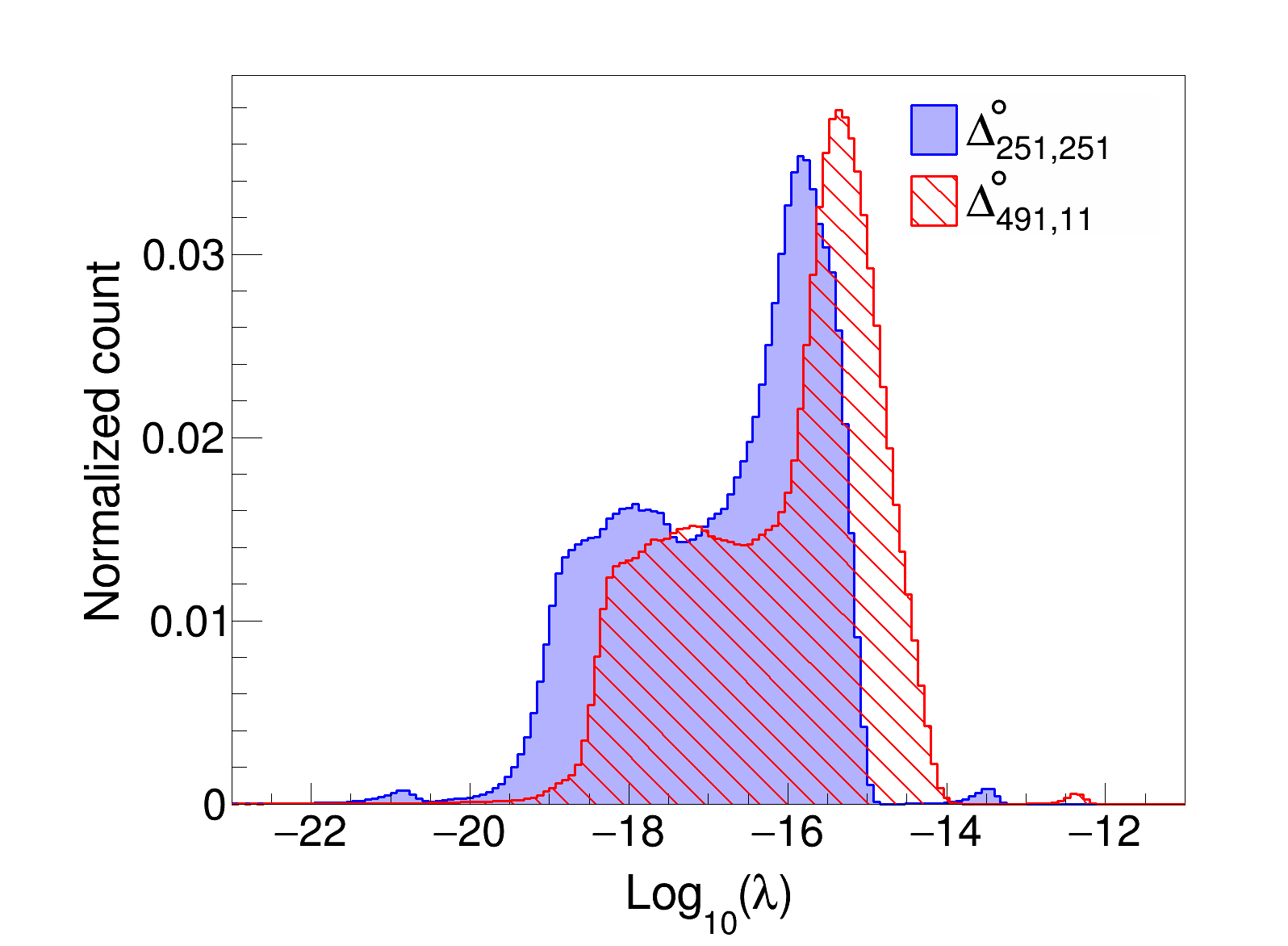}
  \caption{Eigenvalue spectrum of $K_{ij}$}
  \label{fig:keigmax491}
\end{subfigure}
\begin{subfigure}{.5\textwidth}
  \centering
  \includegraphics[width=\linewidth]{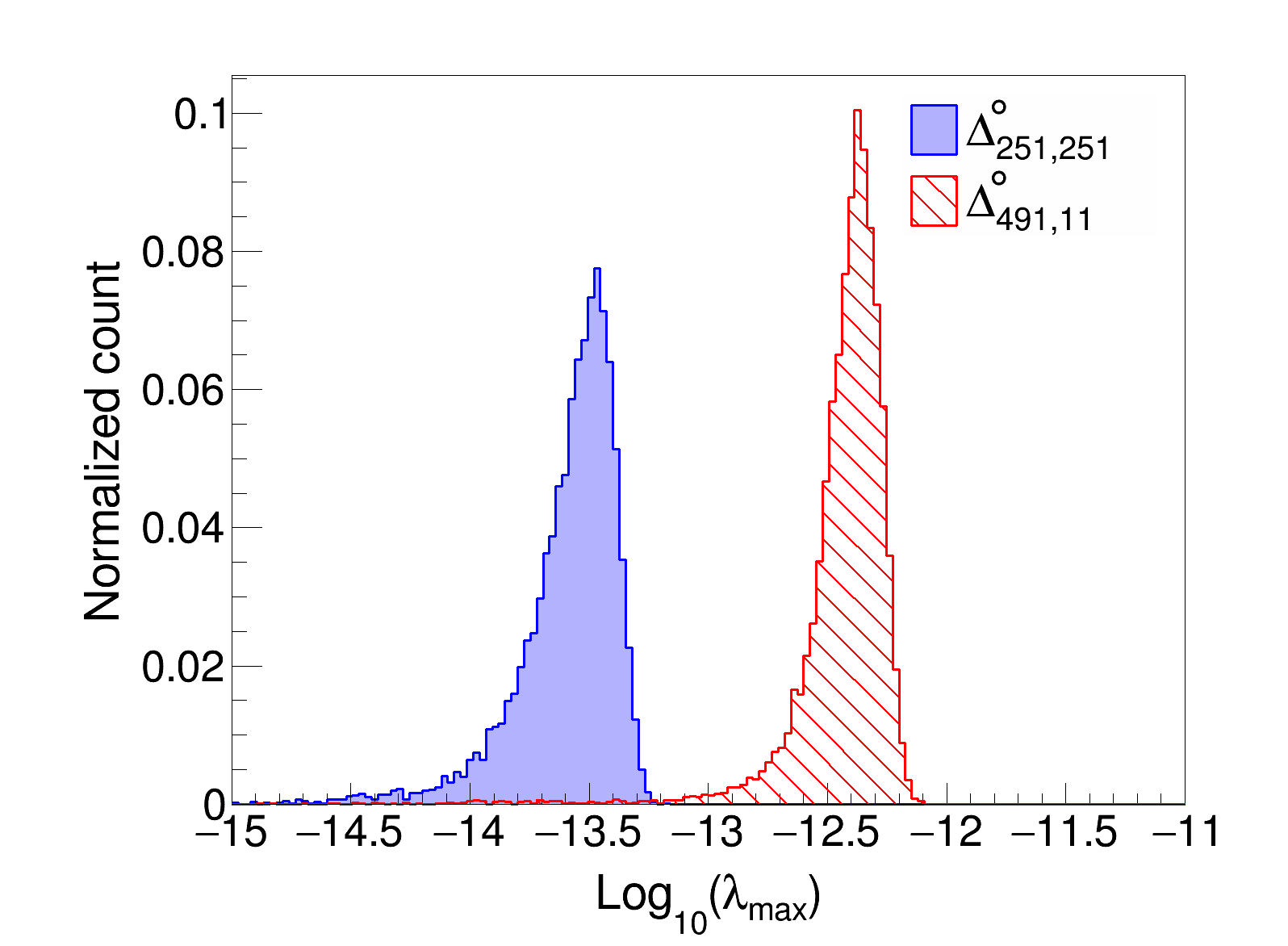}
  \caption{$\lambda_{\mathrm{max}}$, maximum eigenvalue of $K_{ij}$}
  \label{fig:keigmin491}
\end{subfigure}
\caption{Distributions of $d_\text{min}$, cf. \eqref{eq:dmin}, $\cV_{J_\star}$, cf. \eqref{eq:volumes1}, \eqref{eq:strkcone}, \eqref{eq:strkconetip}, and the eigenvalues of $K_{ij}$, cf. \eqref{eq:Kahler}, from a sample of 10,000 triangulations of $\Delta^\circ_{491,11}$ and $\Delta^\circ_{251,251}$, obtained using Algorithm~\ref{algo:random_3} with $N_\text{flip}=40$ and $N_\text{walk}=50$.}
\label{fig:dminvol}
\end{figure}

For $\Delta^\circ_{491,11}$ and $\Delta^\circ_{251,251}$, we use Algorithm \ref{algo:random_3} with $N_\text{flip}=40$ and $N_\text{walk}=50$ to obtain a sample of $10{,}000$ FRSTs for each polytope. We find that each of these FRSTs corresponds to a distinct\footnote{For present purposes, we say that two Calabi-Yau hypersurfaces are distinct if they have different triple intersection numbers $\kappa_{ABC}$. It is in principle possible that two sets of intersection numbers that appear different are related by a basis transformation and the corresponding Calabi-Yau manifolds are equivalent. We do not expect this subtlety to affect any conclusions we draw in this work, though it would matter for an attempt to exactly count, rather than to give an upper bound on, inequivalent threefolds.}
Calabi-Yau hypersurface. The polytope with the largest $h^{2,1}$, $\Delta^\circ_{11,491}$, has 252 FRSTs, all of which give the same Calabi-Yau. For each Calabi-Yau hypersurface $X$, we compute the generators of the Mori cone $\cM_V$, the intersection numbers $\kappa_{ABC}$, the tip of the stretched K\"ahler cone $J_\star$, and the distance $d_\text{min}$.  We then calculate the volume of $X$ and the eigenvalues $\lambda_{\mathrm{min}} \le \cdots \le \lambda_{\mathrm{max}}$ of the K\"ahler metric $K_{ij}$ with $J=J_\star$.  (See \eqref{eq:intnums}, \eqref{eq:strkconetip}, \eqref{eq:dmin}, \eqref{eq:volumes1}, and \eqref{eq:Kahler} for the definitions of these quantities.)

Focusing on the ensembles obtained from $\Delta^\circ_{491,11}$ and $\Delta^\circ_{251,251}$, we first examine the properties of the intersection numbers $\kappa_{ABC}$. For each geometry, we compute the number and root mean square (RMS) size of the nonvanishing intersection numbers, see Figure \ref{fig:intnum}. In both ensembles, the intersection numbers are extremely sparse and the distribution of the RMS size of the nonvanishing entries is strongly peaked.

\begin{table}
\centering
\begin{tabular}{c|c}
    \toprule
    $\mathcal{N}_{\neq0}(\kappa_{ABC})$ & 94\\
    $\mathrm{RMS}(\kappa_{ABC})$  &  62.31\\
    $d_\text{min}$ &  28.86\\
    $\cV_{J_\star}$ &  $9.83 \times 10^{4}$ \\
    $\cV_\text{min}$ &  $9.83 \times 10^{4}$ \\
    $\lambda_\text{min}$ &  $3.42\times 10^{-10}$\\
    $\lambda_\text{max}$ &  $2.34\times10^{-4}$\\
    \bottomrule
\end{tabular}
\caption{The results for the Calabi-Yau hypersurface from $\Delta^\circ_{11,491}$.}
\label{tab:h21_491_results}
\end{table}

The distance $d_\text{min}$, given in Figure \ref{fig:dmin_251-491}, is a good measure of the size of the K\"ahler cones $\cK_V$.\footnote{See \cite{Demirtas:2018akl} for a detailed discussion.} Large values of $d_\text{min}$ indicate that the K\"ahler cones are very narrow. As a direct consequence, the volume of $X$ at $J_\star$, denoted $\cV_{J_\star}$, is large in both ensembles, while the eigenvalues of $K_{ij}$, suppressed by $\cV_{J_\star}$, are accordingly small, see Figure \ref{fig:dminvol}. It is worth noting that although the hypersurfaces from $\Delta^\circ_{491,11}$ have narrower K\"ahler cones compared to that of $\Delta^\circ_{251,251}$, they have smaller volumes $\cV_{J_\star}$.

Finally, we study the Calabi-Yau from $\Delta^\circ_{11,491}$ in more detail. In particular, we construct $\cK_\cup$, cf. \eqref{eq:kcupcone}, and define $J_\star$ as the tip of $\widetilde{\cK_\cup}$, cf. \eqref{eq:kcupcone}. In addition to the quantities discussed above, we numerically minimize $\cV$ in $\widetilde{\cK_\cup}$ and find that it is minimized at $J_\star$. The results are summarized in Table~\ref{tab:h21_491_results}.

\section{Machine Learning Triangulations}\label{sec:ml}

Despite the significant computational advances summarized in \S\ref{sec:software}, constructing a Calabi-Yau hypersurface with $h^{1,1}=491$ and computing the intersection numbers $\kappa_{ABC}$ still takes of order a second on a modern CPU \cite{Cytools}. It is therefore of great interest to design machine learning algorithms to more rapidly predict these topological quantities.  In this section, we will demonstrate that given a training set consisting of a large number of triangulations of a reflexive polytope, a neural network can learn to predict a wide range of geometric quantities with high precision, within $\sim 50\,\mu s$ on a modern GPU.

The main difficulty in making use of a machine learning algorithm in this setting is that the input data (i.e.~an FRST) is not a simple vector in $\mathbb{R}^n$, but is \textit{structured}. Although machine learning with structured data, such as sets, graphs and triangulations, is an area of active research and there are a number of effective approaches,\footnote{See, for example, \cite{DBLP:journals/corr/abs-1806-01261}.} machine learning triangulations directly remains difficult. We now point out a key fact that makes regular triangulations amenable to machine learning: a regular triangulation $\cT$ can be represented efficiently by the GKZ vector $\varphi(\mathcal{T})$.

\paragraph{Datasets}

We consider triangulations of two reflexive polytopes: a randomly chosen polytope with $h^{1,1}=15$ and $h^{2,1}=42$, which we denote by $\Delta^\circ_{15,42}$, and the largest polytope in the Kreuzer-Skarke database, $\Delta^\circ_{491,11}$. Working with a polytope with moderately large $h^{1,1}$ is useful as the number of FRSTs is large enough to train machine learning algorithms while the computational cost of preparing datasets is minimal, allowing for fast experimentation with modest computational resources. On the other hand, $\Delta^\circ_{491,11}$ gives rise to the largest and most challenging triangulations to study.

We start by constructing two million FRSTs for each polytope by picking random height vectors $\mathbf{h} = \mathbf{h}_\circ + \epsilon$, where $\mathbf{h}_\circ$ gives rise to a Delaunay subdivision and the $\epsilon$ are sampled from a Gaussian distribution. We do not use the random sampling algorithms described in \S\ref{sec:sampling}, as preparing a dataset large enough to train a neural network would be prohibitively expensive. For each FRST, we compute the GKZ vector $\varphi(\mathcal{T})$ and the following geometric quantities:

\begin{enumerate}
    \item $\mathcal{N}_{\neq0}(\kappa_{ABC})$: number of nonzero entries of $\kappa_{ABC}$, cf. \eqref{eq:intnums}.
    \item $\mathrm{RMS}(\kappa_{ABC})$: root mean square size of nonzero entries of $\kappa_{ABC}$.
    \item $\kappa_{111}$.
    \item $d_\text{min}$, cf. \eqref{eq:dmin}.
    \item $\cV_{J_\star}$, the volume of the Calabi-Yau hypersurface at $J_\star$, cf. \eqref{eq:volumes1}, \eqref{eq:strkcone}, \eqref{eq:strkconetip}.
    \item $\lambda_\text{min}$, the smallest eigenvalue of $K_{ij}$ at $J_\star$, cf. \eqref{eq:Kahler}.
    \item $\lambda_\text{max}$, the largest eigenvalue of $K_{ij}$ at $J_\star$.
\end{enumerate}
We then construct seven datasets for each polytope, where an element of a dataset consists of a GKZ vector $\varphi(\mathcal{T})$ as the input data and one of the geometric quantities above as the label. We divide each dataset into a training set ($70 \%$), a validation set ($10 \%$), and a testing set ($20 \%$), making sure that a given Calabi-Yau does not appear in both the training/validation and testing sets.\footnote{Note that different triangulations can give rise to the same Calabi-Yau hypersurface. We compute the intersection numbers $\kappa_{ABC}$ to detect identical Calabi-Yau threefolds.}

The statistics of these datasets are summarized in Table \ref{tab:ml}. Note that the statistics for $\Delta^\circ_{491,11}$ differ from those described in \S\ref{sec:geometric_data}. This indicates that the dataset does not consist of a fair random sample of FRSTs. Ultimately, there is a tradeoff between the quality of the sampling algorithm and the size of the dataset. Here, we use a simple sampling algorithm to obtain a sizable training set. Employing the random sampling Algorithm \ref{algo:random_3} to construct datasets of higher quality is likely to improve the performance of the neural networks; we leave this to future work.

\paragraph{Network Architecture}

The most prominent feature of our neural network architecture is the use of ReZero layers \cite{bachlechner2020rezero}. Denoting the output of a dense layer as $F(x_i)$, where $x_i$ are the inputs to the layer, a ReZero layer implements $x_i + \alpha F(x_i)$, where the parameter $\alpha$ is trainable and initialized to zero at the beginning of the training. This layer allows for training very deep neural networks efficiently --- see \cite{bachlechner2020rezero} for more details.

Our architecture consists of a dense layer followed by 32 ReZero layers, all with width 512 and ReLU activation functions.  To regularize the network, we incorporate a Dropout layer in each ReZero layer, except for the last one. For the datasets containing triangulations of $\Delta^\circ_{15,42}$, we use a dropout rate of $0.1$. For $\Delta^\circ_{491,11}$, we observe that the neural networks are much more prone to overfitting, so we use a dropout rate of $0.9$, as well as both L1 and L2 regularization with magnitude $10^{-6}$. We use Tensorflow \cite{abadi2016tensorflow} to implement and train the networks with a learning rate of $5 \times 10^{-3}$ and a batch size of $2^{13}$, on a single Nvidia RTX 2080 Ti GPU.

\begin{table}
\begin{subtable}{\textwidth}
\centering
\begin{tabular}{c|c|c||c|c}
	\toprule
    & $\mu$ & $\sigma$ & MAE & $\mathrm{R}^2$ \\
    \hline
    $\mathcal{N}_{\neq0}(\kappa_{ABC})$ & 114 & 3.396 & $6.92 \times 10^{-2} $ & 1.00 \\
    $\mathrm{RMS}(\kappa_{ABC})$        & 2.35 & $9.72 \times 10^{-2}$ & $8.08 \times 10^{-3}$ & 1.00 \\
    $\log_{10} d_\text{min}$            & 1.92 & $9.44 \times 10^{-2}$ & $4.07 \times 10^{-3}$ & 0.996 \\
    $\log_{10} \cV_{J_\star}$               & 4.13 & 0.290 & $1.41 \times 10^{-2}$ & 0.995 \\
    $\log_{10} \lambda_\text{min}$      & -6.74 & 0.387 & $1.81 \times 10^{-2}$ & 0.999 \\
    $\log_{10} \lambda_\text{max}$      & -5.23 & 0.316 & $1.82 \times 10^{-2}$ & 0.993 \\
    $\kappa_{111}$                    & 3.80 & 2.34 & $1.32 \times 10^{-2}$ & 1.00 \\
    \bottomrule
\end{tabular}
\caption{$h^{1,1}=15$}
\end{subtable}
\newline
\vspace*{1 cm}
\newline
\begin{subtable}{\textwidth}
\centering
\begin{tabular}{c|c|c||c|c}
	\toprule
    & $\mu$ & $\sigma$ & MAE & $\mathrm{R}^2$ \\
    \hline
    $\mathcal{N}_{\neq0}(\kappa_{ABC})$ & 3906 & 12.8 & 5.62 & 0.698 \\
    $\mathrm{RMS}(\kappa_{ABC})$        & 7.98 & $1.76 \times 10^{-2}$ & $6.94 \times 10^{-3}$ & 0.756 \\
    $\log_{10} d_\text{min}$            & 5.83 & $1.48 \times 10^{-2}$ & $9 \times 10^{-3}$ & 0.403 \\
    $\log_{10} \cV_{J_\star}$               & 13.1 & 0.113 & $2.69 \times 10^{-2}$ & 0.909 \\
    $\log_{10} \lambda_\text{min}$      & -18.5 & 0.156 & $3.54 \times 10^{-2}$ & 0.917 \\
    $\log_{10} \lambda_\text{max}$      & -12.8 & 0.811 & $8.48 \times 10^{-2}$ & 0.980 \\
    $\kappa_{111}$                    & 5.00 & 1.91 & $6.52 \times 10^{-2}$ & 0.999 \\
    \bottomrule
\end{tabular}
\caption{$h^{1,1}=491$}
\end{subtable}

\caption{Mean ($\mu$) and standard deviation ($\sigma$) of the labels and performance of the neural networks given in terms of the mean absolute error (MAE), cf.~\eqref{eq:mae} and the coefficient of determination ($\mathrm{R}^2$), cf.~\eqref{eq:rsquared}.}
\label{tab:ml}
\end{table}

\paragraph{Results}

We quantify the performance of the neural networks via the mean absolute error
\begin{equation}
    \mathrm{MAE} :=\frac{1}{N}\sum_{i=1}^N |y_i - x_i|\,,
    \label{eq:mae}
\end{equation}
and the coefficient of determination,
\begin{equation}
    \mathrm{R}^2 := 1- \frac{\sum_{i=1}^N (y_i - x_i)^2}{\sum_{i=1}^N (y_i - \langle x \rangle)^2}\,,
    \label{eq:rsquared}
\end{equation}
where $N$ is the size of the dataset, $y_i$ are the predictions, $x_i$ are the true values and $\langle x \rangle$ is the mean of the $x_i$. Note that $\mathrm{R}^2$ takes values in $(- \infty, 1]$: a perfect predictor achieves $\mathrm{R}^2=1$ while a naive model that predicts the mean $\langle x \rangle$ every time has $\mathrm{R}^2=0$.

We summarize the performance of the neural networks in Table \ref{tab:ml}. At $h^{1,1}=15$ the networks achieve essentially-perfect performance with $\mathrm{R}^2>0.99$ for all seven quantities. While the results are less consistent at $h^{1,1}=491$, it is worth noting that the networks still achieve very high precision with $\mathrm{R}^2>0.9$ for some quantities most relevant for the phenomenology of string compactifications, such as $\cV_{J_\star}$ and $\lambda_\text{max}$.

\section{Conclusions and Outlook}\label{conclusions}

In order to count and understand a vast class of solutions of string theory, we have surveyed Calabi-Yau threefold hypersurfaces in toric varieties.

We first proved an upper bound $N_{\text{FRST}} < 1.53 \times 10^{928}$ on the number of fine, regular, star triangulations (FRSTs) of reflexive polytopes in the Kreuzer-Skarke list, by means of counting lattice points in the associated secondary polytopes.
We then proved an upper bound, $N_{\text{CY}} < 1.65 \times 10^{428}$, on the number of inequivalent Calabi-Yau threefold hypersurfaces resulting from the Kreuzer-Skarke list.  Both bounds were dominated by the largest polytope in the list, with $h^{1,1}=491$ and $h^{2,1}=11$.

In order to explore these extremely large sets, we then developed algorithms and software to generate representative samples of FRSTs and Calabi-Yau threefolds, including the most challenging --- but also, potentially, the most numerous --- cases, where $h^{1,1} \gg 1$.
Our methods allow computation of the topological data of a Calabi-Yau threefold with $h^{1,1}=491$ in of order a second.

Finally, to more rapidly survey the space of geometries, we devised a machine learning algorithm that can predict the topological data of a Calabi-Yau threefold in $\sim 50\,\mu s$, with nearly perfect accuracy for $h^{1,1} \sim 15$, and with good accuracy for some quantities even at $h^{1,1}=491$.  The key to our approach is representation of the data in terms of the GKZ vector of the secondary polytope construction (see Definition \ref{def:gkz_vector}).
In a time when it is not clear which problems in physics and mathematics are amenable to machine learning, we find it worth emphasizing that neural networks can evidently excel at predicting properties of regular triangulations of lattice polytopes, and the corresponding Calabi-Yau hypersurfaces in toric varieties, provided that the triangulation data is represented by the GKZ vector.

We are preparing a user-friendly Python package for constructing and manipulating Calabi-Yau hypersurfaces in toric varieties, with the capability to carry out the calculations described above \cite{Cytools}.
An obvious next step is to deploy our methods in a systematic study of Calabi-Yau threefold hypersurfaces.

Perhaps the most urgent question left unanswered by our work is a \emph{lower} bound on the number of distinct Calabi-Yau threefolds arising from the Kreuzer-Skarke list, or from individual polytopes in it.  Our upper bounds are dominated by the largest polytope, and it is not unreasonable to guess that there are exponentially many inequivalent Calabi-Yau threefolds with $h^{1,1}=491$, but this is as yet unproven.

More generally, computing the actual sizes of the astronomically large discrete sets in the string landscape is an important open problem.

\section*{Acknowledgements}

We thank Geoffrey Fatin, Naomi Gendler, Jim Halverson, Manki Kim, Cody Long, Jakob Moritz, Mike Stillman, and John Stout for useful discussions,
and we are indebted to Mike Stillman for his insights about computing intersection numbers.  We thank Thomas Bachlechner for drawing our attention to \cite{bachlechner2020rezero}.  This work was supported in part by NSF grant PHY-1719877.

\appendix
\section{Testing the Random Walk Algorithm}  \label{sec:app_randomsample}
In \S\ref{sec:sampling}, we described three algorithms for obtaining random samples of Calabi-Yau hypersurfaces. We now test these algorithms by constructing random samples of FRSTs of a reflexive polytope with $h^{1,1}=11$, defined by the vertices
\begin{equation}
    \Delta^\circ_\textsl{test} =
    \begin{pmatrix}
        v_1 & v_2 & v_3 & v_4 & v_5\\
        \hline
        -1 & 1 & -1 & -1 & -1 \\
        -1 & -1 & -1 & 7 & -1 \\
        0 & 0 & 0 & -2 & 2 \\
        0 & 0 & 2 & -2 & 0 \\
    \end{pmatrix}.
\end{equation}
This polytope has $15$ lattice points not strictly interior to facets and is favorable. It has $330{,}019$ FRSTs, corresponding to $117{,}893$ distinct Calabi-Yau hypersurfaces.

\begin{figure}
\begin{subfigure}{.325\textwidth}
  \centering
  \includegraphics[width=\linewidth]{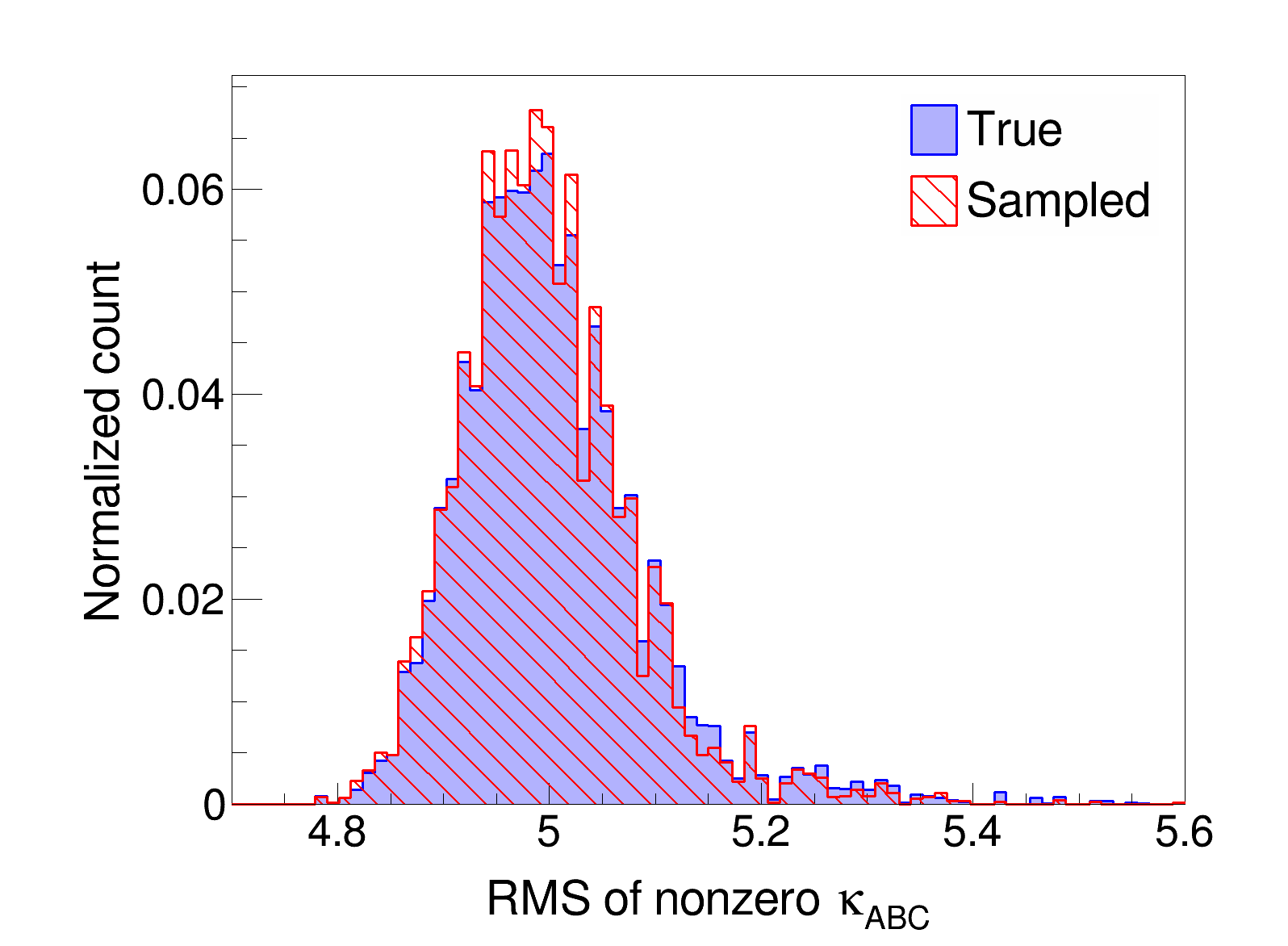}
  \caption{Algorithm \ref{algo:random_1}}
  \label{fig:appintnumrmsalg1}
\end{subfigure}
\begin{subfigure}{.325\textwidth}
  \centering
  \includegraphics[width=\linewidth]{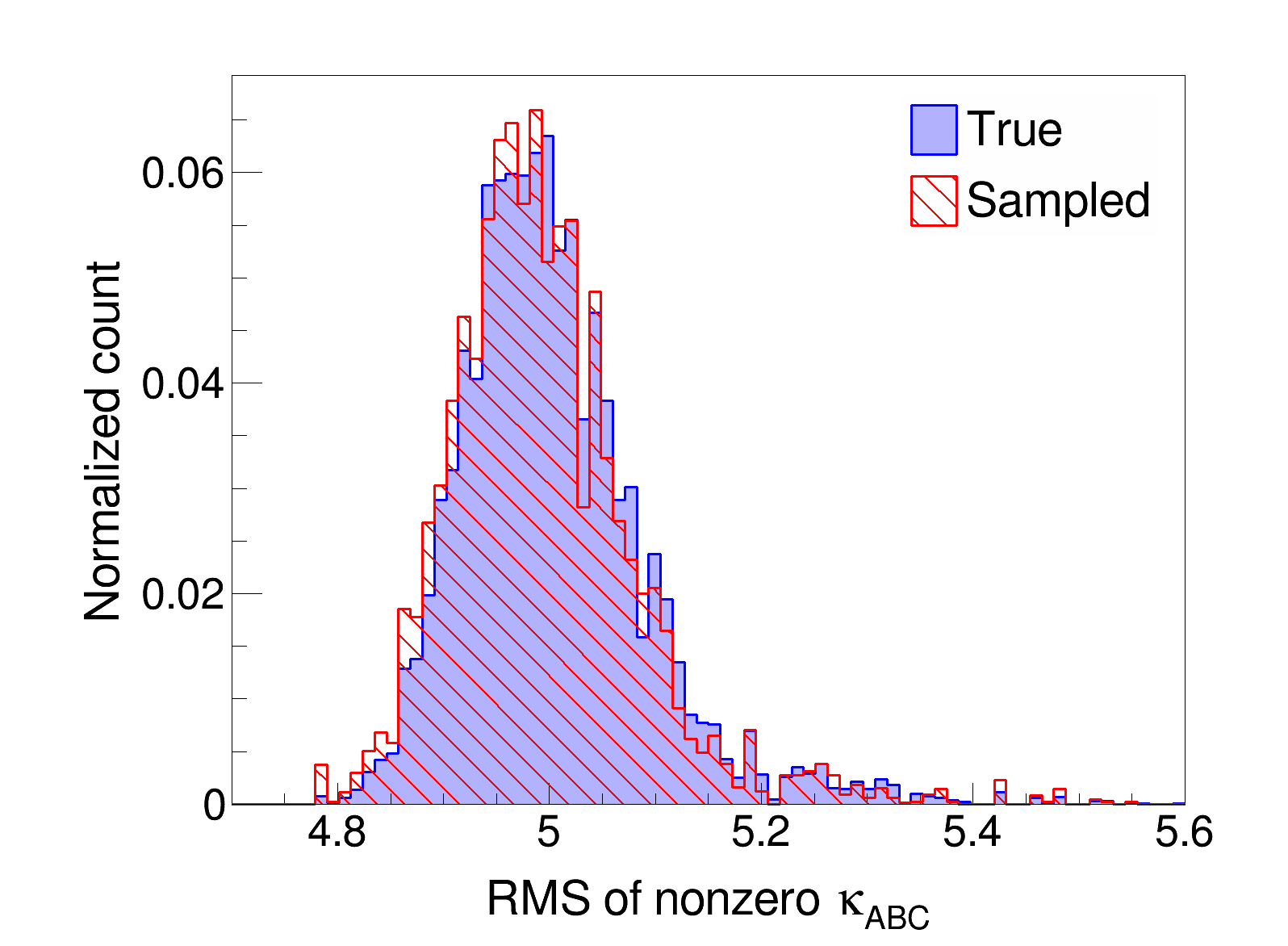}
  \caption{Algorithm \ref{algo:random_2}}
  \label{fig:appintnumrmsalg2}
\end{subfigure}
\begin{subfigure}{.325\textwidth}
  \centering
  \includegraphics[width=\linewidth]{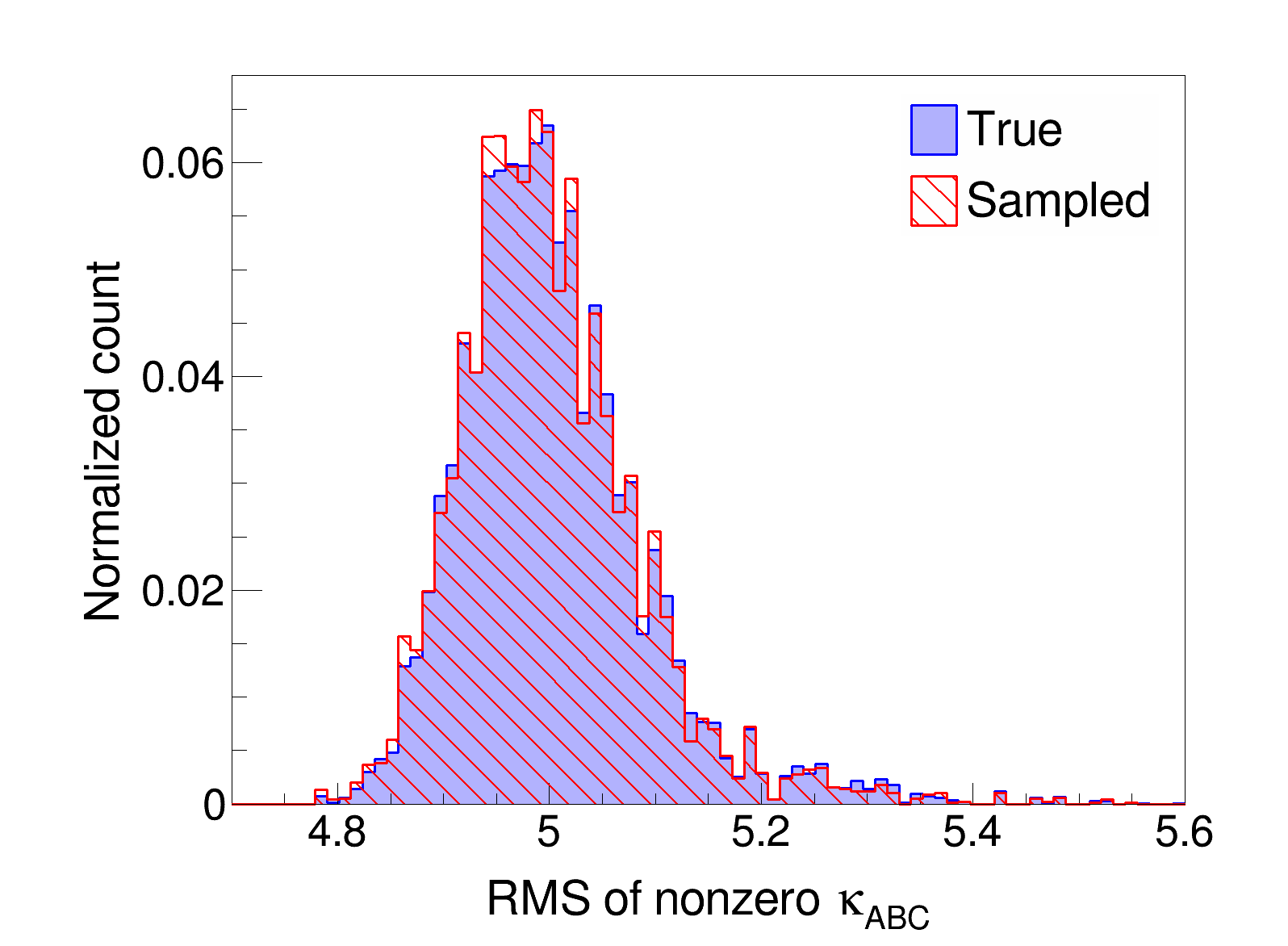}
  \caption{Algorithm \ref{algo:random_3}}
  \label{fig:appintnumrmsalg3}
\end{subfigure}
\caption{Comparison of the root mean square of nonzero intersection numbers from samples obtained using our sampling algorithms with $N_\text{walk}=40$ and $N_\text{step}=50$ versus the true distribution.}
\label{fig:apprms}
\end{figure}

\begin{figure}
\begin{subfigure}{.325\textwidth}
  \centering
  \includegraphics[width=\linewidth]{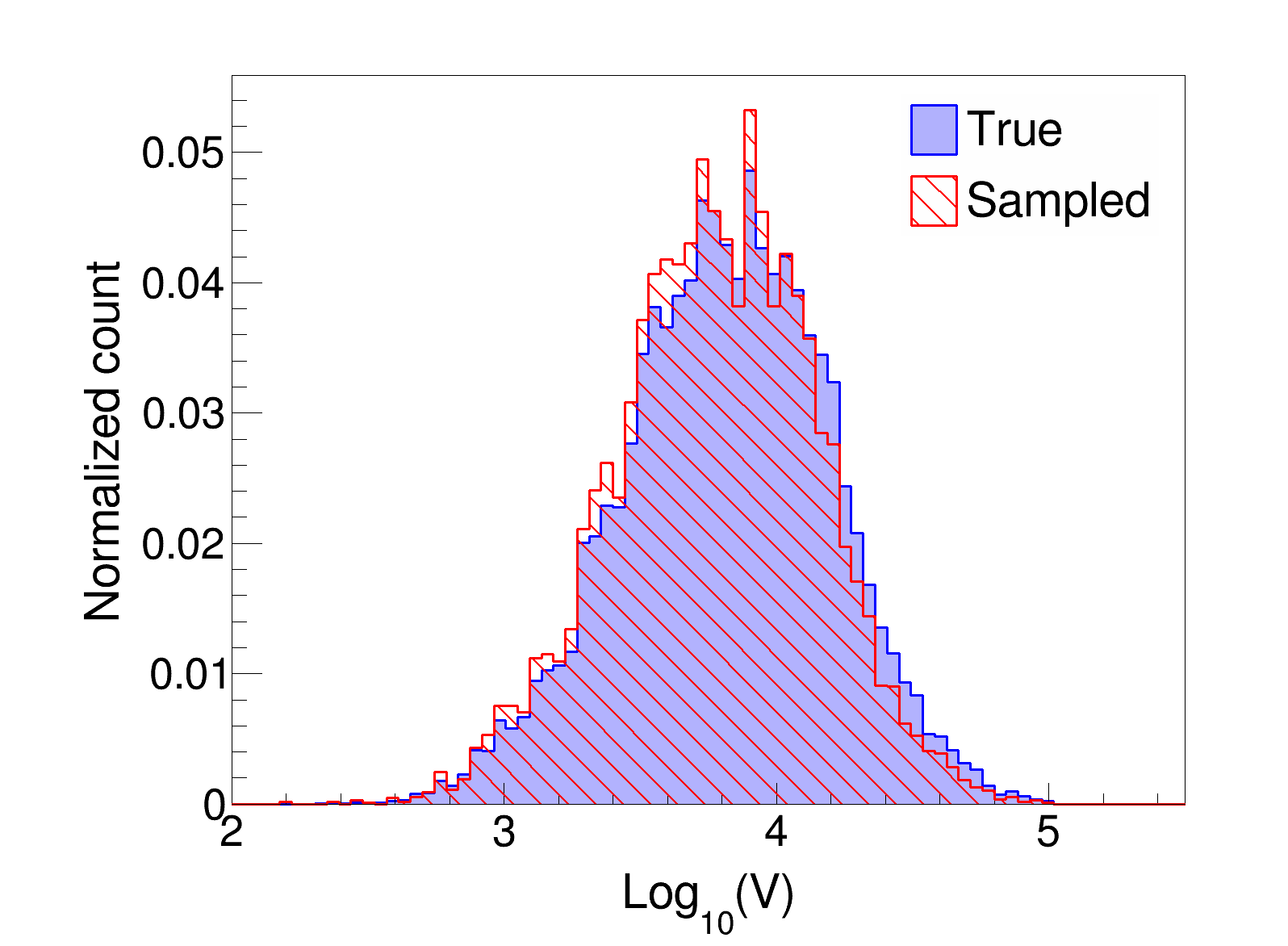}
  \caption{Algorithm \ref{algo:random_1}}
  \label{fig:appcyvolalg1}
\end{subfigure}
\begin{subfigure}{.325\textwidth}
  \centering
  \includegraphics[width=\linewidth]{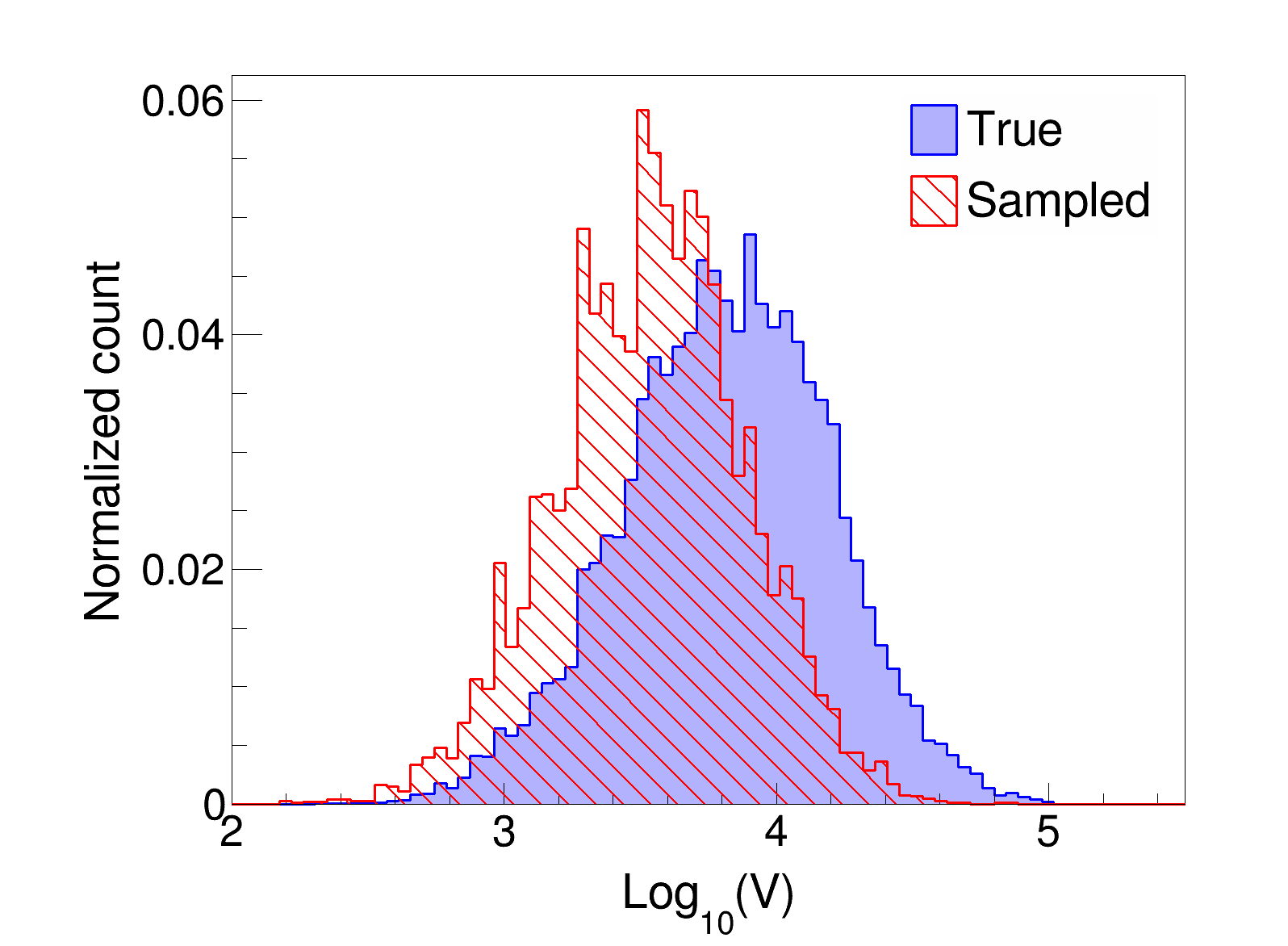}
  \caption{Algorithm \ref{algo:random_2}}
  \label{fig:appcyvolalg2}
\end{subfigure}
\begin{subfigure}{.325\textwidth}
  \centering
  \includegraphics[width=\linewidth]{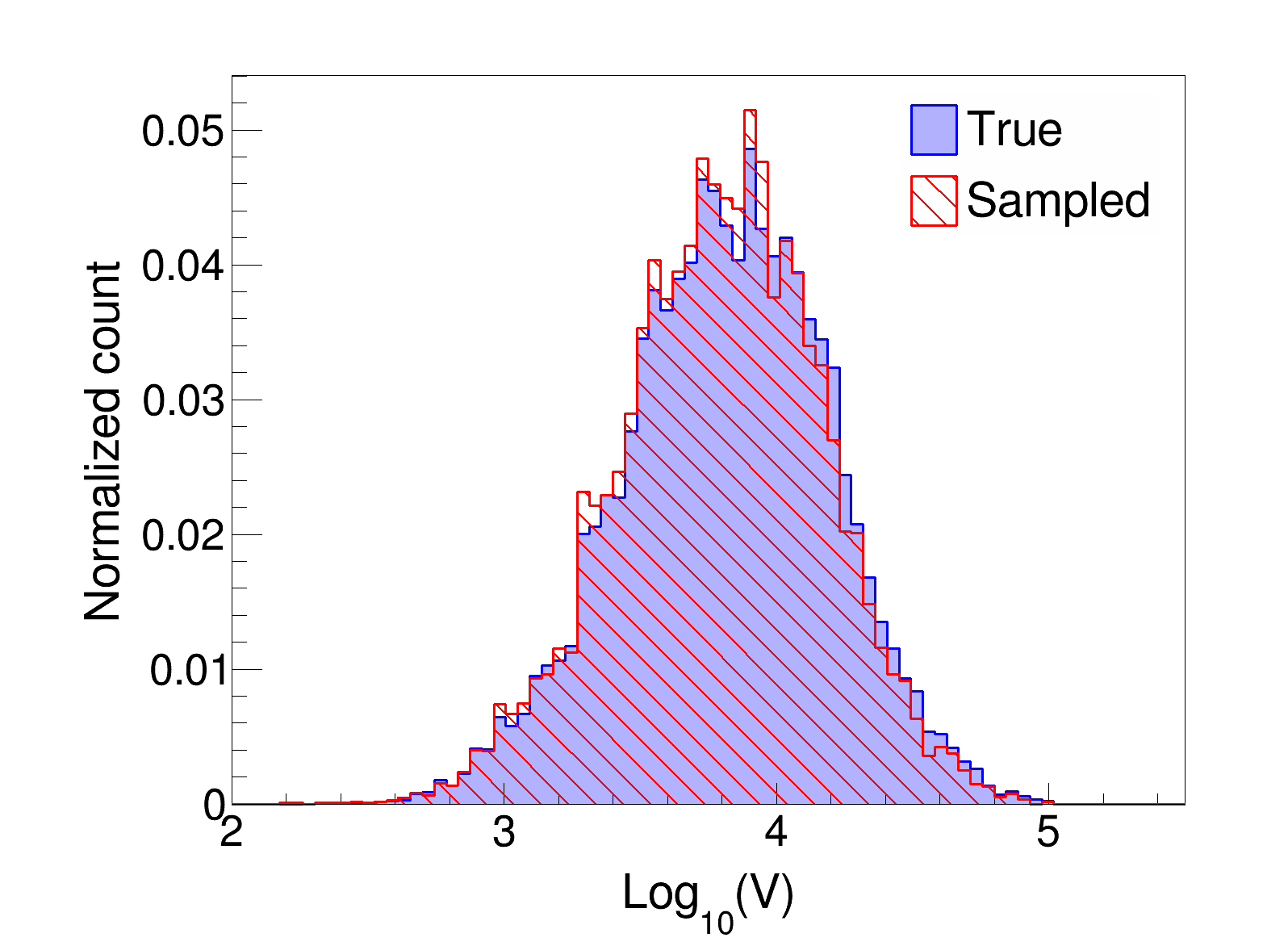}
  \caption{Algorithm \ref{algo:random_3}}
  \label{fig:appcyvolalg3}
\end{subfigure}
\caption{Comparison of the volume of the Calabi-Yau at $J_\star$ from samples obtained using our sampling algorithms with $N_\text{flip}=40$ and $N_\text{walk}=50$ versus the true distribution.}
\label{fig:appvol}
\end{figure}

As we can explicitly construct the complete set of Calabi-Yau hypersurfaces arising from this polytope, we can determine whether a given sample is representative. To this end, we construct three samples of $10{,}000$ FRSTs using Algorithms \ref{algo:random_1}, \ref{algo:random_2} and \ref{algo:random_3}, with $N_\text{walk}=50$ and $N_\text{flip}=40$. All three algorithms return samples for which the properties of the intersection numbers $\kappa_{ABC}$ match the true distribution. The comparison for the root mean square of the nonzero intersection numbers is given in Figure \ref{fig:apprms}. However, Algorithm \ref{algo:random_2} prefers FRSTs with wider K\"ahler cones and, as a result, smaller volumes: see Figure \ref{fig:appvol}. As discussed in \S\ref{sec:sampling}, this behavior is due to the fact that Algorithm \ref{algo:random_2} generates samples where FRSTs corresponding to larger cones in the secondary fan are favored.

\begin{figure}
\begin{subfigure}{.325\textwidth}
  \centering
  \includegraphics[width=\linewidth]{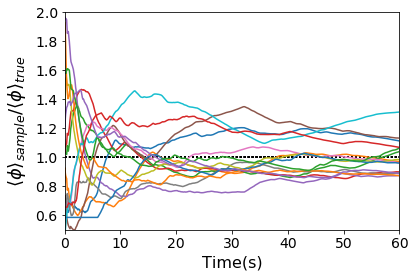}
  \caption{Algorithm \ref{algo:random_1}}
  \label{fig:appcmaalg1}
\end{subfigure}
\begin{subfigure}{.325\textwidth}
  \centering
  \includegraphics[width=\linewidth]{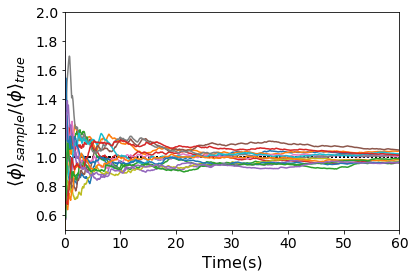}
  \caption{Algorithm \ref{algo:random_2}}
  \label{fig:appcmaalg2}
\end{subfigure}
\begin{subfigure}{.325\textwidth}
  \centering
  \includegraphics[width=\linewidth]{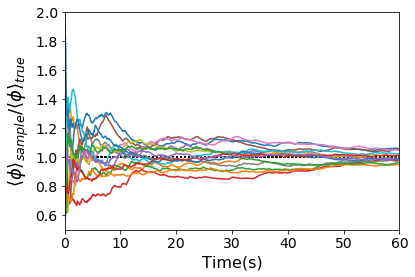}
  \caption{Algorithm \ref{algo:random_3}}
  \label{fig:appcmaalg3}
\end{subfigure}
\caption{ $\langle \varphi^i \rangle_\text{sample} / \langle \varphi^i \rangle_\text{true}$ vs. run-time of Algorithm \ref{algo:random_1} with $N_\text{flip}=1$, Algorithm \ref{algo:random_2} with $N_\text{walk}=8$ and Algorithm 3 with $N_\text{flip}=1$, $N_\text{walk}=4$.}
\label{fig:appcma}
\end{figure}

A random sampling algorithm is useful only if the statistics of the sample it produces converge within a reasonable time frame. We compare the convergence rates of our algorithms by computing the GKZ vector $\varphi^i$, cf. Definition \ref{def:gkz_vector}, for each triangulation in the sample and comparing the mean value $\langle \varphi^i \rangle_\text{sample}$ of the sample to the true mean value $\langle \varphi^i \rangle_\text{true}$. As the algorithm converges, the ratio $\langle \varphi^i \rangle_\text{sample} / \langle \varphi^i \rangle_\text{true}$ approaches 1 for all $i$. In Figure \ref{fig:appcma}, we plot $\langle \varphi^i \rangle_\text{sample} / \langle \varphi^i \rangle_\text{true}$ for Algorithms \ref{algo:random_1}, \ref{algo:random_2} and \ref{algo:random_3} versus run-time on a single CPU, adjusting $N_\text{flip}$ and $N_\text{walk}$ such that all three algorithms produce samples of $\sim 1000$ triangulations per minute. While Algorithms \ref{algo:random_2} and \ref{algo:random_3} converge reasonably quickly, Algorithm \ref{algo:random_1} is significantly slower.

In summary, testing the random sampling algorithms at small $h^{1,1}$ reveals that Algorithm \ref{algo:random_1} converges slowly, Algorithm \ref{algo:random_2} produces skewed samples, and Algorithm \ref{algo:random_3} produces fair samples efficiently.

\newpage

\bibliographystyle{JHEP}
\bibliography{refs}

\providecommand{\href}[2]{#2}\begingroup\raggedright\begin{thebibliography}{10}

\bibitem{Kreuzer:2000xy}
M.~Kreuzer and H.~Skarke, \emph{{Complete classification of reflexive polyhedra
  in four-dimensions}},
  \href{http://dx.doi.org/10.4310/ATMP.2000.v4.n6.a2}{\emph{Adv. Theor. Math.
  Phys.} {\bf 4} (2002) 1209--1230},
  [\href{http://arxiv.org/abs/hep-th/0002240}{{\tt hep-th/0002240}}].

\bibitem{Altman:2014bfa}
R.~Altman, J.~Gray, Y.-H. He, V.~Jejjala and B.~D. Nelson, \emph{{A Calabi-Yau
  Database: Threefolds Constructed from the Kreuzer-Skarke List}},
  \href{http://dx.doi.org/10.1007/JHEP02(2015)158}{\emph{JHEP} {\bf 02} (2015)
  158}, [\href{http://arxiv.org/abs/1411.1418}{{\tt 1411.1418}}].

\bibitem{Huang:2018gpl}
Y.-C. Huang and W.~Taylor, \emph{{Comparing elliptic and toric hypersurface
  Calabi-Yau threefolds at large Hodge numbers}},
  \href{http://dx.doi.org/10.1007/JHEP02(2019)087}{\emph{JHEP} {\bf 02} (2019)
  087}, [\href{http://arxiv.org/abs/1805.05907}{{\tt 1805.05907}}].

\bibitem{2004JHEP...07..069A}
B.~Allanach, D.~Grellscheid and F.~Quevedo, \emph{{Genetic algorithms and
  experimental discrimination of SUSY models}},
  \href{http://dx.doi.org/10.1088/1126-6708/2004/07/069}{\emph{JHEP} {\bf 07}
  (2004) 069}, [\href{http://arxiv.org/abs/hep-ph/0406277}{{\tt
  hep-ph/0406277}}].

\bibitem{2014JHEP...08..010A}
S.~Abel and J.~Rizos, \emph{{Genetic Algorithms and the Search for Viable
  String Vacua}}, \href{http://dx.doi.org/10.1007/JHEP08(2014)010}{\emph{JHEP}
  {\bf 08} (2014) 010}, [\href{http://arxiv.org/abs/1404.7359}{{\tt
  1404.7359}}].

\bibitem{2016JHEP...03..045C}
M.~Cirafici, \emph{{Persistent Homology and String Vacua}},
  \href{http://dx.doi.org/10.1007/JHEP03(2016)045}{\emph{JHEP} {\bf 03} (2016)
  045}, [\href{http://arxiv.org/abs/1512.01170}{{\tt 1512.01170}}].

\bibitem{2017PhLB..774..564H}
Y.-H. {He}, \emph{{Machine-learning the string landscape}},
  \href{http://dx.doi.org/10.1016/j.physletb.2017.10.024}{\emph{Physics Letters
  B} {\bf 774} (Nov., 2017) 564--568},
  [\href{http://arxiv.org/abs/1706.02714}{{\tt 1706.02714}}].

\bibitem{Krefl:2017yox}
D.~Krefl and R.-K. Seong, \emph{{Machine Learning of Calabi-Yau Volumes}},
  \href{http://dx.doi.org/10.1103/PhysRevD.96.066014}{\emph{Phys. Rev. D} {\bf
  96} (2017) 066014}, [\href{http://arxiv.org/abs/1706.03346}{{\tt
  1706.03346}}].

\bibitem{2017JHEP...08..038R}
F.~Ruehle, \emph{{Evolving neural networks with genetic algorithms to study the
  String Landscape}},
  \href{http://dx.doi.org/10.1007/JHEP08(2017)038}{\emph{JHEP} {\bf 08} (2017)
  038}, [\href{http://arxiv.org/abs/1706.07024}{{\tt 1706.07024}}].

\bibitem{2017JHEP...09..157C}
J.~Carifio, J.~Halverson, D.~Krioukov and B.~D. Nelson, \emph{{Machine Learning
  in the String Landscape}},
  \href{http://dx.doi.org/10.1007/JHEP09(2017)157}{\emph{JHEP} {\bf 09} (2017)
  157}, [\href{http://arxiv.org/abs/1707.00655}{{\tt 1707.00655}}].

\bibitem{2018arXiv180503615A}
S.~Abel, D.~G. Cerdeño and S.~Robles, \emph{{The Power of Genetic Algorithms:
  what remains of the pMSSM?}},  \href{http://arxiv.org/abs/1805.03615}{{\tt
  1805.03615}}.

\bibitem{2018PhLB..785...65B}
K.~Bull, Y.-H. He, V.~Jejjala and C.~Mishra, \emph{{Machine Learning CICY
  Threefolds}},
  \href{http://dx.doi.org/10.1016/j.physletb.2018.08.008}{\emph{Phys. Lett. B}
  {\bf 785} (2018) 65--72}, [\href{http://arxiv.org/abs/1806.03121}{{\tt
  1806.03121}}].

\bibitem{2019ForPh..67l0084C}
A.~Constantin and A.~Lukas, \emph{{Formulae for Line Bundle Cohomology on
  Calabi-Yau Threefolds}},
  \href{http://dx.doi.org/10.1002/prop.201900084}{\emph{Fortsch. Phys.} {\bf
  67} (2019) 1900084}, [\href{http://arxiv.org/abs/1808.09992}{{\tt
  1808.09992}}].

\bibitem{2018arXiv180902547K}
D.~Klaewer and L.~Schlechter, \emph{{Machine Learning Line Bundle Cohomologies
  of Hypersurfaces in Toric Varieties}},
  \href{http://dx.doi.org/10.1016/j.physletb.2019.01.002}{\emph{Phys. Lett. B}
  {\bf 789} (2019) 438--443}, [\href{http://arxiv.org/abs/1809.02547}{{\tt
  1809.02547}}].

\bibitem{2018arXiv180902612E}
H.~Erbin and S.~Krippendorf, \emph{{GANs for generating EFT models}},
  \href{http://arxiv.org/abs/1809.02612}{{\tt 1809.02612}}.

\bibitem{2019NuPhB.940..113M}
A.~Mütter, E.~Parr and P.~K. Vaudrevange, \emph{{Deep learning in the
  heterotic orbifold landscape}},
  \href{http://dx.doi.org/10.1016/j.nuclphysb.2019.01.013}{\emph{Nucl. Phys. B}
  {\bf 940} (2019) 113--129}, [\href{http://arxiv.org/abs/1811.05993}{{\tt
  1811.05993}}].

\bibitem{2018arXiv181206960C}
A.~Cole and G.~Shiu, \emph{{Topological Data Analysis for the String
  Landscape}}, \href{http://dx.doi.org/10.1007/JHEP03(2019)054}{\emph{JHEP}
  {\bf 03} (2019) 054}, [\href{http://arxiv.org/abs/1812.06960}{{\tt
  1812.06960}}].

\bibitem{2019PhLB..795..700B}
K.~Bull, Y.-H. He, V.~Jejjala and C.~Mishra, \emph{{Getting CICY High}},
  \href{http://dx.doi.org/10.1016/j.physletb.2019.06.067}{\emph{Phys. Lett. B}
  {\bf 795} (2019) 700--706}, [\href{http://arxiv.org/abs/1903.03113}{{\tt
  1903.03113}}].

\bibitem{2019JHEP...06..003H}
J.~Halverson, B.~Nelson and F.~Ruehle, \emph{{Branes with Brains: Exploring
  String Vacua with Deep Reinforcement Learning}},
  \href{http://dx.doi.org/10.1007/JHEP06(2019)003}{\emph{JHEP} {\bf 06} (2019)
  003}, [\href{http://arxiv.org/abs/1903.11616}{{\tt 1903.11616}}].

\bibitem{2019PhLB..79834889H}
Y.-H. He and S.-J. Lee, \emph{{Distinguishing elliptic fibrations with AI}},
  \href{http://dx.doi.org/10.1016/j.physletb.2019.134889}{\emph{Phys. Lett. B}
  {\bf 798} (2019) 134889}, [\href{http://arxiv.org/abs/1904.08530}{{\tt
  1904.08530}}].

\bibitem{2019arXiv191008605A}
A.~Ashmore, Y.-H. He and B.~A. Ovrut, \emph{{Machine learning Calabi-Yau
  metrics}},  \href{http://arxiv.org/abs/1910.08605}{{\tt 1910.08605}}.

\bibitem{2020ForPh..68e0005H}
J.~Halverson and C.~Long, \emph{{Statistical Predictions in String Theory and
  Deep Generative Models}},
  \href{http://dx.doi.org/10.1002/prop.202000005}{\emph{Fortsch. Phys.} {\bf
  68} (2020) 2000005}, [\href{http://arxiv.org/abs/2001.00555}{{\tt
  2001.00555}}].

\bibitem{2020arXiv200313339D}
R.~Deen, Y.-H. He, S.-J. Lee and A.~Lukas, \emph{{Machine Learning String
  Standard Models}},  \href{http://arxiv.org/abs/2003.13339}{{\tt 2003.13339}}.

\bibitem{2020arXiv200616619H}
Y.-H. He and S.-T. Yau, \emph{{Graph Laplacians, Riemannian Manifolds and their
  Machine-Learning}},  \href{http://arxiv.org/abs/2006.16619}{{\tt
  2006.16619}}.

\bibitem{2020arXiv200700009B}
M.~Bies, M.~Cveti\v{c}, R.~Donagi, L.~Lin, M.~Liu and F.~Ruehle, \emph{{Machine
  Learning and Algebraic Approaches towards Complete Matter Spectra in 4d
  F-theory}},  \href{http://arxiv.org/abs/2007.00009}{{\tt 2007.00009}}.

\bibitem{RUEHLE20201}
F.~Ruehle, \emph{{Data science applications to string theory}},
  \href{http://dx.doi.org/10.1016/j.physrep.2019.09.005}{\emph{Phys. Rept.}
  {\bf 839} (2020) 1--117}.

\bibitem{2018arXiv181202893H}
Y.-H. He, \emph{{The Calabi-Yau Landscape: from Geometry, to Physics, to
  Machine-Learning}},  \href{http://arxiv.org/abs/1812.02893}{{\tt
  1812.02893}}.

\bibitem{Gelfand1994}
I.~M. Gelfand, M.~M. Kapranov and A.~V. Zelevinsky, \emph{Discriminants,
  Resultants, and Multidimensional Determinants}.
\newblock Birkh{\"a}user Boston, Boston, MA, 1994,
  \href{http://dx.doi.org/10.1007/978-0-8176-4771-1}{10.1007/978-0-8176-4771-1}.

\bibitem{Batyrev:1994hm}
V.~V. Batyrev, \emph{{Dual polyhedra and mirror symmetry for Calabi-Yau
  hypersurfaces in toric varieties}}, {\emph{J. Alg. Geom.} {\bf 3} (1994)
  493--545}, [\href{http://arxiv.org/abs/alg-geom/9310003}{{\tt
  alg-geom/9310003}}].

\bibitem{Braun:2017nhi}
A.~P. Braun, C.~Long, L.~McAllister, M.~Stillman and B.~Sung, \emph{{The Hodge
  Numbers of Divisors of Calabi-Yau Threefold Hypersurfaces}},
  \href{http://arxiv.org/abs/1712.04946}{{\tt 1712.04946}}.

\bibitem{Gelfand1989}
I.~M. Gelfand, A.~V. Zelevinsky and M.~M. Kapranov, \emph{Newton polyhedra of
  principal a-determinants}, {\emph{Soviet Mathematics. Doklady} {\bf 40}
  (1989) 278--281}.

\bibitem{Gelfand1991}
I.~M. Gelfand, A.~V. Zelevinsky and M.~M. Kapranov, \emph{{Discriminants of
  polynomials in several variables and triangulations of Newton polyhedra}},
  {\emph{Leningrad Math. J.} {\bf 2} (1991) 499--505}.

\bibitem{DeLoera2010}
J.~A. De~Loera, J.~Rambau and F.~Santos, \emph{Triangulations: Structures for
  Algorithms and Applications}.
\newblock Springer Berlin Heidelberg, Berlin, Heidelberg, 2010,
  \href{http://dx.doi.org/10.1007/978-3-642-12971-1}{10.1007/978-3-642-12971-1}.

\bibitem{oda1991}
T.~Oda and H.~S. Park, \emph{{Linear Gale transforms and
  Gel'fand-Kapranov-Zelevinskij decompositions}},
  \href{http://dx.doi.org/10.2748/tmj/1178227461}{\emph{Tohoku Math. J. (2)}
  {\bf 43} (1991) 375--399}.

\bibitem{Anclin2003}
E.~E. Anclin, \emph{An upper bound for the number of planar lattice
  triangulations},
  \href{http://dx.doi.org/https://doi.org/10.1016/S0097-3165(03)00097-9}{\emph{Journal
  of Combinatorial Theory, Series A} {\bf 103} (2003) 383 -- 386}.

\bibitem{Kaibel03}
V.~Kaibel and G.~M. Ziegler, \emph{Counting lattice triangulations},  in
  \emph{Surveys in Combinatorics 2003, number 307 in Lond. Math. Soc. Lect.
  Note Ser}, pp.~277--308, Cambridge University Press, 2003.
\newblock \href{http://arxiv.org/abs/math/0211268}{{\tt math/0211268}}.

\bibitem{Altman:2018zlc}
R.~Altman, J.~Carifio, J.~Halverson and B.~D. Nelson, \emph{{Estimating
  Calabi-Yau Hypersurface and Triangulation Counts with Equation Learners}},
  \href{http://dx.doi.org/10.1007/JHEP03(2019)186}{\emph{JHEP} {\bf 03} (2019)
  186}, [\href{http://arxiv.org/abs/1811.06490}{{\tt 1811.06490}}].

\bibitem{Wall1966}
C.~T.~C. Wall, \emph{{Classification problems in differential topology. V}},
  \href{http://dx.doi.org/10.1007/BF01389738}{\emph{Inventiones mathematicae}
  {\bf 1} (Dec, 1966) 355--374}.

\bibitem{Hubsch:1992nu}
T.~Hubsch, \emph{{Calabi-Yau manifolds: A Bestiary for physicists}}.
\newblock World Scientific, Singapore, 1994.

\bibitem{Rambau2002}
J.~Rambau, \emph{{TOPCOM}: Triangulations of point configurations and oriented
  matroids},  in \emph{Mathematical Software---ICMS 2002} (A.~M. Cohen, X.-S.
  Gao and N.~Takayama, eds.), pp.~330--340, World Scientific, 2002.
\newblock
  \href{http://dx.doi.org/https://doi.org/10.1142/9789812777171_0035}{DOI}.

\bibitem{caputo2015}
P.~Caputo, F.~Martinelli, A.~Sinclair and A.~Stauffer, \emph{Random lattice
  triangulations: Structure and algorithms},
  \href{http://dx.doi.org/10.1214/14-AAP1033}{\emph{Ann. Appl. Probab.} {\bf
  25} (06, 2015) 1650--1685}.

\bibitem{Demirtas:2018akl}
M.~Demirtas, C.~Long, L.~McAllister and M.~Stillman, \emph{{The Kreuzer-Skarke
  Axiverse}}, \href{http://dx.doi.org/10.1007/JHEP04(2020)138}{\emph{JHEP} {\bf
  04} (2020) 138}, [\href{http://arxiv.org/abs/1808.01282}{{\tt 1808.01282}}].

\bibitem{Kreuzer_2004}
M.~Kreuzer and H.~Skarke, \emph{Palp: A package for analysing lattice polytopes
  with applications to toric geometry},
  \href{http://dx.doi.org/10.1016/s0010-4655(03)00491-0}{\emph{Computer Physics
  Communications} {\bf 157} (Feb, 2004) 87–106}.

\bibitem{BAGNARA20083}
R.~Bagnara, P.~M. Hill and E.~Zaffanella, \emph{The parma polyhedra library:
  Toward a complete set of numerical abstractions for the analysis and
  verification of hardware and software systems},
  \href{http://dx.doi.org/https://doi.org/10.1016/j.scico.2007.08.001}{\emph{Science
  of Computer Programming} {\bf 72} (2008) 3 -- 21}. Special issue on
  experimental software and toolkits (EST).

\bibitem{sagemath}
{The Sage Developers}, \emph{{S}ageMath, the {S}age {M}athematics {S}oftware
  {S}ystem ({V}ersion 9.0)}, 2020.

\bibitem{Barber96}
C.~B. Barber, D.~P. Dobkin and H.~Huhdanpaa, \emph{The quickhull algorithm for
  convex hulls}, {\emph{{ACM} Transactions on Mathematical Software} {\bf 22}
  (1996) 469--483}.

\bibitem{Braun:2014xka}
A.~P. Braun and T.~Watari, \emph{{The Vertical, the Horizontal and the Rest:
  anatomy of the middle cohomology of Calabi-Yau fourfolds and F-theory
  applications}}, \href{http://dx.doi.org/10.1007/JHEP01(2015)047}{\emph{JHEP}
  {\bf 01} (2015) 047}, [\href{http://arxiv.org/abs/1408.6167}{{\tt
  1408.6167}}].

\bibitem{cgal:eb-19a}
{The CGAL Project}, \emph{{CGAL} User and Reference Manual}.
\newblock {CGAL Editorial Board}, {4.14}~ed., 2019.

\bibitem{cgal:dDTriangs}
O.~Devillers, S.~Hornus and C.~Jamin, \emph{{dD} triangulations},  in
  \emph{{CGAL} User and Reference Manual}.
\newblock {CGAL Editorial Board}, {4.14}~ed., 2019.

\bibitem{Jordan2018}
C.~Jordan, M.~Joswig and L.~Kastner, \emph{Parallel enumeration of
  triangulations}, {\emph{Electronic Journal of Combinatorics} {\bf 25} (2018)
  3.6}, [\href{http://arxiv.org/abs/1709.04746}{{\tt 1709.04746}}].

\bibitem{ortools}
L.~Perron and V.~Furnon, \emph{{OR-Tools, Version 7.4}},
  \url{https://developers.google.com/optimization/}.

\bibitem{Berglund1995}
P.~Berglund, S.~H. Katz and A.~Klemm, \emph{{Mirror symmetry and the moduli
  space for generic hypersurfaces in toric varieties}},
  \href{http://dx.doi.org/10.1016/0550-3213(95)00434-2}{\emph{Nucl. Phys. B}
  {\bf 456} (1995) 153--204}, [\href{http://arxiv.org/abs/hep-th/9506091}{{\tt
  hep-th/9506091}}].

\bibitem{mosek}
{MOSEK ApS}, \emph{Mosek {Optimizer} {API} for {Python}},
  \url{http://docs.mosek.com/9.1/pythonapi/index.html}.

\bibitem{Mathematica}
{Wolfram Research{,} Inc.}, \emph{{Mathematica, {V}ersion 12.0}},
  \url{https://www.wolfram.com/mathematica}.

\bibitem{10.1145/1391989.1391995}
Y.~Chen, T.~A. Davis, W.~W. Hager and S.~Rajamanickam, \emph{Algorithm 887:
  Cholmod, supernodal sparse cholesky factorization and update/downdate},
  \href{http://dx.doi.org/10.1145/1391989.1391995}{\emph{ACM Trans. Math.
  Softw.} {\bf 35} (Oct., 2008) }.

\bibitem{Cytools}
M.~Demirtas, L.~McAllister, A.~Rios-Tascon et~al., \emph{{work in progress}}.

\bibitem{DBLP:journals/corr/abs-1806-01261}
P.~W. Battaglia, J.~B. Hamrick, V.~Bapst, A.~Sanchez{-}Gonzalez, V.~F.
  Zambaldi, M.~Malinowski et~al., \emph{Relational inductive biases, deep
  learning, and graph networks}, {\emph{CoRR} (2018) },
  [\href{http://arxiv.org/abs/1806.01261}{{\tt 1806.01261}}].

\bibitem{bachlechner2020rezero}
T.~Bachlechner, B.~P. Majumder, H.~H. Mao, G.~W. Cottrell and J.~McAuley,
  \emph{Rezero is all you need: Fast convergence at large depth},
  \href{http://arxiv.org/abs/2003.04887}{{\tt 2003.04887}}.

\bibitem{abadi2016tensorflow}
M.~Abadi, P.~Barham, J.~Chen, Z.~Chen, A.~Davis, J.~Dean et~al.,
  \emph{Tensorflow: A system for large-scale machine learning},  in \emph{12th
  {USENIX} Symposium on Operating Systems Design and Implementation ({OSDI
  16)}}, 2016.

\end{thebibliography}\endgroup

\end{document}